\documentclass{article}

\usepackage{arxiv}

\usepackage[utf8]{inputenc} 
\usepackage[T1]{fontenc}    
\usepackage{hyperref}       
\usepackage{url}            
\usepackage{booktabs}       
\usepackage{amsfonts}       
\usepackage{nicefrac}       
\usepackage{microtype}      
\usepackage{lipsum}
\usepackage{graphicx}
\usepackage{listings}
\usepackage{multirow}

\graphicspath{ {./images/} }

\newtheorem{definition}{CSPm Definition}

\title{Groovy Parallel Patterns – A Process Oriented Parallelization Library}

\author{
 Jon Kerridge \\
  School of Coumputing \\
  Edinburgh Napier University\\
  10 Colinton Road\\
  Edinburgh EH10 5DT UK\\
  \texttt{j.kerridge@napier.ac.uk} \\
  orcid=0000-0002-1261-9519\\
   \And
 Neil Urquhart \\
  School of Coumputing \\
  Edinburgh Napier University\\
  10 Colinton Road\\
  Edinburgh EH10 5DT UK\\
  \texttt{n.urquhart@napier.ac.uk} \\
  orcid=0000-0001-5669-5479\\

}

\begin{document}
\maketitle
\begin{abstract}
A novel parallel patterns library, Groovy Parallel Patterns, is presented which, from the outset, has been designed to exploit more general process parallelism than the usual data and task parallel architectures.  The library executes on a standard Java Virtual Machine.  The library provides a collection of processes that can be plugged together to form a variety of parallel architectures and is intrinsically its own DSL.  A network of processes is guaranteed to be deadlock and livelock free and terminate correctly and this is proved by the use of formal methods.  Error capture and a basic logging mechanism have been incorporated.  The library enables effective refinement of solutions between process networks which can be checked also using formal methods. A library user is only required to create the required methods as pieces of sequential code, typically taken from extant sequential solutions, which can then be invoked by the processes as required.  The utility of the library is demonstrated by several examples including; Monte Carlo Methods, Concordance, Jacobi solutions, N-body problems and Mandelbrot, which is implemented on both a multicore processor and a workstation cluster.  The examples are analysed for speedup and efficiency, which show good and consistent performance improvement up to the number of available processor cores and workstations.

\end{abstract}


\section{Introduction}
The concept of Parallel Programming Skeletons has been widely explored since Cole published his original manifesto in 2004 \cite{cole_2004}. In the intervening period, many academic groups have developed skeleton based libraries for a variety of implementation environments \cite{gonzalez_2010}.   The concept of structured parallel programming based on a set of building blocks has been proposed by Danelutto and Torquati \cite{danelutto_torquati_2013}.  Even though the academic Computing community has been active \cite{alg_skeleton} there has been little uptake of these concepts by the wider computing community, although in the period October 2014 to December 2019 a book \cite{kerridge_2014} explaining the effective use of concurrency and parallelism has been downloaded more than 0.8 million times (Bookboon Private Communication).  It should be noted that, the concept of a generally applicable process-based parallel farming architecture appeared in 1988 \cite{Allen-1998}.
It is difficult to determine why the wider computing community finds concurrency and parallelism difficult.  It could be down to education, the underlying thread models and low-level communication mechanisms used, the way current skeleton models force the programmer to think in a skeleton specific manner or the fact that people think it is hard to transform initial program designs into concurrent and parallel solutions.  The latter of these is probably the biggest obstacle to the adoption of concurrent and parallel programming ideas, together with many programmers believing they must implement some form of thread based solution \cite{lea_2000}.  The concept that concurrency and parallelism is difficult is re-enforced by examples such as, the Introductory chapter of \cite{mskenney}  includes the statement “However, even though parallel programming might not be as hard as is commonly advertised, it is often more work than is sequential programming”. This paper argues this need not be the case, provided more appropriate process based data flow style components are used when transforming an initial design into a parallel solution.  This style of parallel design can take advantage of the ideas and theoretical developments associated with Communicating Sequential Processes \cite{roscoe_2010,Hoare-1978} and its associated machine readable language CSPm \cite{gibson-robinson_armstrong_boulgakov_roscoe} together with the checking tool FDR \cite{gibson-robinson_armstrong_boulgakov_roscoe_2014}.  
Groovy Parallel Patterns (GPP)   comprises a library of processes that can be combined into data flow and other architectures that solve problems using communicating processes without  the need for a deep theoretical understanding of the underlying concepts or the need to understand the parallel architecture upon which the solution is to be deployed, apart from realising that a multi-core solution and a cluster based solution may have some minor differences.  A further advantage of GPP is that the depth of understanding of parallel solutions required to make the first steps is small, when compared with most of the other skeleton solutions.  The fundamental GPP concept is that we can take a data flow diagram of the required solution and transform it into a set of processes together with some understanding that partitioning data is a common first step in designing parallel solutions.  A further requirement was that the solution could be executed as a pure sequential solution without change to the underlying methods, even if they had to be invoked in a somewhat different manner, thereby ensuring that the basic solution is known to function correctly before it is parallelised.

The collected contribution of the GPP Library is the integration of:
\begin{itemize}
    \item Extended collection of parallel building blocks, both functional and connector,
\item Formal proof of the correct operation of process networks,
\item Logging integrated from the outset,
\item DSL builder makes system development easier using a declarative network definition,
\item Use of extant sequential code expected and supported,
\item Exploit parallel processing easily and as efficiently as possible,
\item Cluster based solutions requiring no change to Library or User code, except for network invocation.
\end{itemize}

The paper is structured as follows; Section 2 describes the technology used in the construction of the GPP library and shows that much of it is already available and thus the theoretical underpinning is well understood and long established, including the ability to generate formal proofs of the correctness that can be automatically checked.  Section 3 provides a motivating example, based on Montecarlo methods, to demonstrate the ease and brevity with which a parallel architecture can be constructed using GPP.  It highlights the unique aspect of the library which enables a sequential invocation of the algorithm as well as one using a parallel architecture. It demonstrates how the structure of the library provides its own DSL (Domain Specific Language).  Section 4 provides a description of the basic data objects used to provide the declarative style of the library, while Section 5 introduces the higher-level components of the library.  Section 6 illustrates how the GPP library can be used to solve some architecturally common problems, showing that the solutions are reliably scalable in the number of available processing cores.  Section 7 then shows how the architecture can be applied to a workstation cluster, with similarly predictable and uniform performance improvements, provided the underlying algorithm has such properties.  Section 8 describes the logging system built into the library so that performance bottlenecks can be identified, and its use is explained with reference to the concordance example.  Section 9 shows how the theoretical underpinning of the library can be used to reason about the behaviour of process networks and enables the construction of different architectures which, though they produce the same result, may have different performance characteristics.  In section 10, the GPP library is compared with some of the more common parallelisation libraries to identify the benefits it provides.  Section 11 provides an evaluation of the Library.  Finally, in Sections 12 and 13 some conclusions are drawn and areas for further research identified.
\section{Review of Background and Available Technology}
The GPP library was initially under development concurrently with the approach described in \cite{danelutto_torquati_2013}.  The main idea was to have a set of small building blocks from which larger systems could be constructed to build common architectural patterns, such as Farms and Pipelines, used in parallel processing.  The authors chose to use a thread-based implementation mechanism.  GPP chose to implement similar concepts using the higher-level process abstraction.  Some of the concepts have had to be modified slightly but the basic dataflow architecture has been retained.  Further, GPP has augmented the set of basic building blocks with components that allow multi-core access to shared data objects used, for example, in large matrix and image processing applications.
A proof of concept attempt at building a process-based implementation of the concepts was described in \cite{chalmers_jon_pederson_2016}, in which all the processes and the overall process networks were constructed by hand.  There was no attempt to prove, using an automated tool such as FDR, that the components would work as expected, nor was the idea of using the application network as a declarative script for a DSL to build the application.  Further, the application code and the process bodies were not as well separated as they are in GPP because there was no attempt to take extant sequential code and use it with as few changes as possible in a parallel solution. Finally, the ability to log the interactions in the process network so that bottlenecks can be isolated was not even considered, which is now included in GPP.  In particular, the ability for the user to specify the object property that is to be logged as objects are passed from one process to the next is novel.
In the following sections, the technical basis of the GPP library is placed in context and how it exploits Groovy \cite{groovy_programming_language19} to make application of the library simpler, building upon extant and well proven technologies.

\subsection{Communicating Process Architectures}

Hoare’s Communicating Sequential Processes (CSP) \cite{Hoare-1978} provides the theoretical underpinning for the GPP library.  The Occam language \cite{inmos_ltd_1984} initially implemented the CSP concepts and this lead to the Laws of Occam \cite{roscoe_hoare_1988} which allow the refinement of Occam programs to show the equivalence of different program structures.  The Java Communicating Sequential Processes library (JCSP) \cite{cspforjava}, directly implemented the Occam model of parallelism, comprising processes and communication channels, in standard Java. JCSP is implemented in standard Java using the native Thread, Synchronisation and Monitor model. This complexity is hidden from the JCSP user. Processes encapsulate all data and typically comprise a repeated sequence of code which communicates with other processes by copying data from one process to another.  A communication channel provides a unidirectional, unbuffered and synchronised transfer capability between processes.  Thus, one process writes to a channel and another reads from that channel.  Whichever process attempts to communicate first, waits, idle until the other process is ready at which point the data is copied from the writing process to the reading process.  An idle process consumes no processing resource whatsoever.  Once the communication is complete both processes can continue in parallel.  No data is shared between processes.  In JCSP object references are copied between processes and thus an additional requirement is imposed such that once an object has been transferred to another process the original process can no longer refer to that object.  In JCSP this requirement must be enforced by the programmer, but in the case of GPP this aspect is guaranteed by the design of the processes.    The underlying JCSP process and communication method has a formal proof of its operational correctness \cite{welch_martin_2000Java, welch_martin_2000CSP} thus we can be confident that the GPP library is built on sound principles that enable the programmer to be confident that the underlying model is correct.

The design of GPP utilises a number of formal design patterns that ensure a process network utilising these patterns is guaranteed to be deadlock free.  Welch et al \cite{welch_justo_wilcock_1993} showed that two patterns known as I/O-PAR and I/O-SEQ provided this capability.  Both patterns assume that a code sequence is to be repeatedly undertaken.  All the processes in GPP conform to the I/O SEQ pattern (see Section 9).  The correctness of this approach is further proved using CSPm models of the processes which are checked using FDR4.  A further pattern called Client-Server \cite{hansen_1973}, is also used in cluster-based solutions (see section \ref{sec:workstations}).

The GPP library has been designed using language, process, communication and pattern primitives that have all been subject to formal derivation or have a theoretical underpinning. This is not the case with the skeleton libraries discussed in Section 10. Those based on underlying thread models are very difficult to check formally.  Users of the GPP Library can be confident that once a system has been built the parallel parts of the system will behave as expected.
\subsection{Groovy Support}
Apache Groovy \cite{groovy_programming_language19} is an object-oriented dynamic language for the Java platform, it aims to improve productivity by using an easy to learn syntax. Groovy \cite{Konig_2015} support for JCSP was first described in  \cite{kerridge_barclay_savage_2005} and was introduced to make the concurrent and parallel computing aspects much clearer, mainly by removing the need for much of the boiler-plate code Java requires.  This was further developed in \cite{kerridge_2014} which showed how parallel systems could be easily developed for multi-core and cluster-based systems. A set of helper classes have been developed as a Groovy Library called, groovyJCSP \cite{kerridge_18,kerridge_2014}, to make use of JCSP even simpler.  GroovyJCSP comprises four helper classes; PAR that cause a list of processes to run in parallel, ALT that provides a non-deterministic choice over a list of inputs, and Channel Input and Output Lists that allow the manipulation of Lists (arrays) of channels.  These channels can be either internal memory-based channels for communications between processes on the same processor or networked channels when communicating between workstations in a cluster.  In JCSP a process definition is independent of the channel type and thus there is no difference to the process definition if it involves internal or networked channel communications. The drawback of this approach was that a programmer, using their own process and communication networks, had to be fully conversant with the design patterns and underlying JCSP requirements to effect solutions.  GPP was thus developed to remove this requirement for the novice parallel programmer and which could exploit some of the more recent Groovy language enhancements.  This is further exploited by means of a builder program which removes the need for the programmer to construct the process network and the required channel declarations, effectively providing a DSL  for the construction of parallel systems.  The programmer defines the objects which contain the sequential application code for the problem together with some definitional objects that are required by the GPP processes.  The process network is specified as a sequence of process definitions through which the application objects flow.  The builder program then transforms the declarative style into runnable Groovy.    

Groovy does provide concurrent programming support (\cite{Konig_2015}, Chapter18) using the Actor model \cite{Hewitt73} in a package called GPpars.  The Actor model does have some associated laws that inform its use \cite{Hewitt1977}, similar to \cite{roscoe_hoare_1988} but does not have the automated support provided by FDR \cite{gibson-robinson_armstrong_boulgakov_roscoe} for CSP based systems.  Systems built using GPars do have a similar dataflow architecture to that created using GPP but the Actor model operates at a less abstract level where the user has to consider the messages that are passed between the Actors.  It would not be possible to prove the correct operation of a GPars model as we can with GPP.

\section{Motivating Example: Montecarlo $\pi$}

\begin{figure}[h]
        \centering
        \includegraphics[width=100mm]{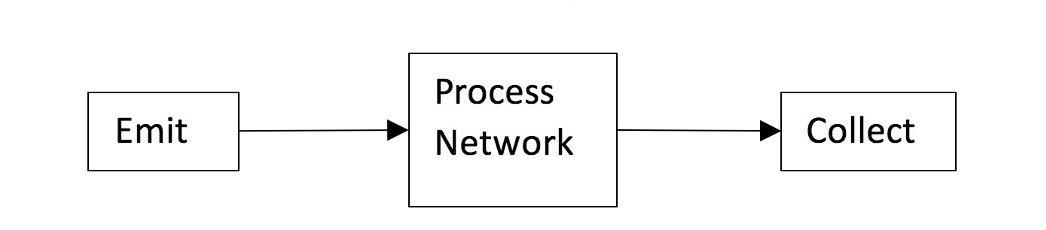}
        \caption{Fundamental Dataflow Pattern}
        \label{fig:new1}
    \end{figure}

The fundamental pattern used by GPP is shown in figure \ref{fig:new1}.  The Emit process will send many data objects to the Process Network.  The Process Network undertakes the application process.  Once the Process Network has completed the processing of an object it will send it to the Collect process where the results can be processed and collated as required. In some cases, the application can be achieved using an even simpler parallel architecture such as a data parallel farm or a task parallel pipeline, making the deployment even easier because the Emit and Collect processes are contained in the pattern as in Data Parallel (Farm) or Task Parallel (Pipeline) applications.

\subsection{The GPP Library Code}

The value of $\pi$ can be calculated using Montecarlo methods.  Such methods are commonly used in parallel solutions to simulation problems \cite{monte_carlo}.  The algorithm presented here requires the generation of many random points lying within the upper unit quadrant.  The ratio of points that lie within the unit radius quadrant to those without is $\pi/4$.  Thus, the requirement is to create as many points as possible and determine if they lie within or out with the unit quadrant. Using a multi-core processor, we can partition this determination over each of the cores, thereby speeding up performance or increasing the accuracy of the solution

Solutions are specified in a declarative style using specialised objects that define the data objects to be processed and the results to be collected.  It is assumed that the initial process will emit many data objects   into the network and these will be processed as and when resources become available.  The results are obtained by a specialised collection process.  This follows a pattern described in (\cite{kerridge_2014} pp. (ii) 147-166).  The required Groovy data objects are shown in Listing \ref{lst:listing1}, where $piData$ \{Listing \ref{lst:listing1}:4-8\} and $piResults$ \{Listing \ref{lst:listing1}:10-13\} are the Groovy object definitions the user has written.

\begin{minipage}{\textwidth}
\begin{lstlisting}[caption = Data Detail Object Definitions for Montecarlo $\pi$,label=lst:listing1]
int workers = 4                 // number of parallel worker processes
int instances = 1024            // number of object instances to create, 2048, 4096
int iterations = 100000            // number of random points per instance
def emitData = new DataDetails ( dName: piData.getName(),
                                 dInitMethod: piData.init,
                                 dInitData: [instances],
                                 dCreateMethod: piData.create,
                                 dCreateData: [iterations])

def resultDetails = new ResultDetails ( rName: piResults.getName(),
                                        rInitMethod: piResults.init,
                                        rCollectMethod: piResults.collector,
                                        rFinaliseMethod: piResults.finalise)


\end{lstlisting}
\end{minipage}

The value of \textit{workers} \{Listing \ref{lst:listing1}:1\} is the number of processor cores to be used, $instances$ gives the number of objects to be emitted into the network and $iterations$ \{Listing \ref{lst:listing1}:3\} the number of random points each object instance is to evaluate.  This subdivision of the task would be second nature to a confident parallel programmer but not so to a purely sequential programmer.  One of the design goals of the library is to make it easier to appreciate such design decisions.

The object $piData$ (Listing \{\ref{lst:listing1}:4-8\} ) has methods whose names are held as $String$ constants in the properties $init$, $create$ and $withinOp$. The object $piResults$ (\{Listing \ref{lst:listing1}:10-13\}) similarly has methods identified by $init$, $collector$ and $finalise$.

\begin{minipage}{\textwidth}
\begin{lstlisting}[caption = Invoking the $DataParallelCollect$ Pattern for Montecarlo $\pi$,label=lst:listing2]
def piFarm = new DataParallelCollect ( eDetails: emitData,
                                       rDetails: resultDetails,
                                       workers : workers,
                                       function: piData.withinOp )

\end{lstlisting}
\end{minipage}

These $DataDetails$ objects can then be used by processes within the GPP library, as shown in Listing \ref{lst:listing2} where they are used to invoke a data parallel architecture or farm built into the library.  The $gppBuilder$ application then takes the combination of Listing \ref{lst:listing1} and Listing \ref{lst:listing2} and converts them to runnable Groovy code that will distribute the work over 4 $workers$ (Listing \{\ref{lst:listing2}:3\}) cores.  The farm will receive 1024 ($instances$) objects of type $piData$ each of which evaluates 100000 ($iterations$) random points.

\begin{figure}[h]
        \centering
        \includegraphics[width=100mm]{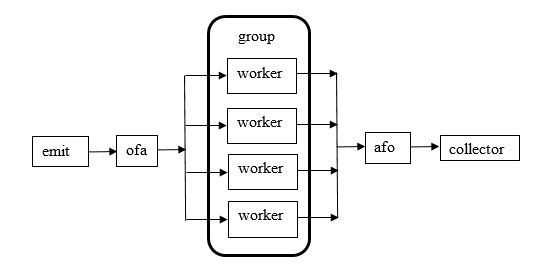}
        \caption{Process Structure Created by Listing \ref{lst:listing3}}
        \label{fig:3-1}
    \end{figure}

The $DataParallelCollect$ pattern simply needs to know the $DataDetails$ \{Listing \ref{lst:listing2}:1,2\} object that defines how data is emitted into the network, $eDetails$ and how the subsequent results are collected, $rDetails$.  The pattern will invoke $workers$ parallel processes each of which will undertake the operation named as $function$ \{Listing \ref{lst:listing2}:4\}.  Listing \ref{lst:listing3} shows how the same process structure can be created using lower-level components of the library and is represented diagrammatically in Figure \ref{fig:3-1}. Thus, by combining the content of Listing \ref{lst:listing1} and Listing \ref{lst:listing3} the library has the same effect as the combination of Listing \ref{lst:listing1} and Listing \ref{lst:listing2} .  It should also be noted that the network definition given in Listing \ref{lst:listing3} contains no communication channel declarations.  The $gppBuilder$ program creates the required channel definitions and adds them to the process definitions automatically so the user is not concerned with the detail of channel creation.  Similarly, the user does not need to create the parallel invocation of the processes.

\begin{minipage}{\textwidth}
\begin{lstlisting}[caption = The Effective Internal Process Structure of the $DataParallelCollect$ Pattern,label=lst:listing3]
def emit = new Emit (eDetails: emitData)
def ofa = new OneFanAny (destinations: workers)
def group = new AnyGroupAny (workers: workers, function: piData.withinOp)
def afo = new AnyFanOne (sources: workers)
def collector = new Collect (rDetails: resultDetails)

\end{lstlisting}
\end{minipage}

The $init$ \{Listing \ref{lst:listing1}:5,11\} methods  are used to initialise the data object classes.  In the case of $piData$ \{Listing \ref{lst:listing1}:7\} the number of object instances to be created is specified. The $init$ method of $piResults$ does nothing.  The create method of $piData$ is used to initialise the number of iterations to be undertaken by each object as it passes through the process network.  The $piData$ method, referred to as $withinOp$ \{Listing \ref{lst:listing3}:3\}, is used to determine the number of points that lie within the unit circle from the iterations calculations performed.  The collector method of $piResults$ adds the number of points that are within the unit circle to a running total of all the data object instances.  The finalise method of $piResults$ then determines the value of $\pi$ using the ratio of the total points that were within the unit circle to the total number of points that were evaluated ($instances * iterations$).

Figure \ref{fig:3-1} shows both the individual processes that form the data flow architecture together with the internal structure of the group process.  Figure \ref{fig:3-1} is very similar, apart from names, to that given in \cite{Allen-1998} but with the addition of the \textit{OneFanAny (ofa)} and \textit{AnyFanOne (afo)} processes. The processes $Emit$ and $Collect$ are examples of $terminal$ processes.  Processes \textit{ofa} and \textit{afo} are examples of $connector$ processes and $AnyGroupAny (aga)$ of the $functional$ processes. Process $aga$ is in fact a parallel collection of $workers$ $Worker$ processes.  A property of the $gppBuilder$ program is that it takes the above declarations and, if it can construct a legal network, then it is guaranteed to be deadlock and livelock free and will also terminate and recover all resources used by the network.  Of course, it may not work correctly if the methods written by the user do not work as expected.  However, the user can take the objects that are used within the parallel network and invoke them in a purely sequential manner to ensure they function as required.  Listing \ref{lst:listing4} shows a suitable sequential code that exercises the objects.

\begin{minipage}{\textwidth}
\begin{lstlisting}[caption = Sequential Invocation of the MCpi Objects ,label=lst:listing4]
int instances = 1024
int iterations = 100000
def mcpires = new piResults()
for ( i in 1 .. instances){
    def mcpi = new piData()
    mcpi.initClass([instances])
    mcpi.createInstance([iterations])
    mcpi.getWithin(null)
    mcpires.collector(mcpi)
}
mcpires.finalise(null)
\end{lstlisting}
\end{minipage}

In Listing \ref{lst:listing4} the object methods are called in the usual manner rather than using a $String$ representation of the name of the method.  Regardless of the method of invocation, the user must construct two objects that implement the required methods referred to in Listing \ref{lst:listing1} to Listing \ref{lst:listing4} for $piData$ and $piResults$ and these are briefly discussed in the next sub-sections.
\subsubsection{The $piData$ Object}
Listing \ref{lst:listing5} gives the $piData$ definition.  All methods return an integer value that is used by the calling process to determine the method outcome.  Typically, the return value is $completedOK$   \{Listing \ref{lst:listing5}:12\} unless an error occurs in which case the method must return a negative value.  The $createInstance$ method \{Listing \ref{lst:listing5}:15-23\} is the only one that has different return values.  This method is called in an $Emit$ terminal type process.  The $Emit$ process needs to know if the last object instance has been created or not; the values $normalTermination$ \{Listing \ref{lst:listing5}:16\} and $normalContinuation$  \{Listing \ref{lst:listing5}:21\} are used to indicate these states respectively.  Most of the user written methods have a $List$ parameter that enables the passing of a variable number of parameters as required by the method and the application.  In some cases, the parameters must be specifically ordered and formatted.  User methods never return a value other than $completedOK$ or an error value.

The effect of $getWithin()$ \{Listing \ref{lst:listing5}:25-34\} is to calculate the number of points within the unit quadrant for the object instance being processed. The final determination of the value of $\pi$ can only be determined once all the object instances have been processed. In Listings \ref{lst:listing2} and \ref{lst:listing3} the $getWithin()$ method is referred to by its name $withinOp$  \{Listing \ref{lst:listing2}:4,\ref{lst:listing3}:3\}.

\begin{minipage}{\textwidth}
\begin{lstlisting}[caption = The $piData$ object definition,label=lst:listing5]
class piData extends gpp.DataClass {
  int iterations = 0                      // used to record the number of iterations
  int within = 0                          // counts the number of points within quadrant
  static int instance = 1                 // records the emitted object instance number
  static int instances                    // total number of instances to be created
  static String withinOp = "getWithin"    // exported names of object methods
  static String init = "initClass"        // exported names do not have to match actual
  static String create = "createInstance"

  int initClass (List p) {
    instances = p[0]                      // initialise instances
    return completedOK
  }

  int createInstance (List d){
    if ( instance > instances) return normalTermination  //all instances created; terminate
    else {
      iterations = d[0]            // initialise the number of iterations
      within = 0                   // initialise within
      instance = instance + 1      // increment the instance number
      return normalContinuation    // more instances to create; continue
    }
  }

  int getWithin(List d){
    def rng = new Random()
    float x, y
    for ( i in 1 ..iterations){
      x = rng.nextFloat()
      y = rng.nextFloat()
      if ( ((x*x) + (y*y)) <= 1.0 ) within = within + 1
    }
    return completedOK
  }
}
\end{lstlisting}
\end{minipage}

\subsubsection{The $piResults$ object}

Listing \ref{lst:listing6} shows the definition of the $piResults$ class. The $collector$ method (Listing \{\ref{lst:listing6}:11-15\}) has a specific parameter ($o$), which is the object that has been read by the Collector process.  The object type is not typed in the parameter list but can be explicitly stated in the method body to ensure it is of type $piData$.  The parameter $o$ is supplied by the $Collect$ process, as described in the documentation \cite{kerridge_19a} as is the requirements for all the other required methods.  The method simply accumulates the $within$ values, as well as the total number of $iterations$.  The $finalise$ \{Listing \ref{lst:listing6}:17-20\} method computes and prints the value of $\pi$.

\begin{minipage}[c]{0.95\textwidth}
\begin{lstlisting}[caption = The $piResults$ object definition,label=lst:listing6]
class piResults extends gpp.DataClass {
  int iterationSum = 0                    //holds total number of iterations
  int withinSum = 0                       //holds total points within quadrant
  static String init = "initClass"        //exported names of object methods
  static String collector = "collector"
  static String finalise = "finalise"

  int initClass (List d){
    return completedOK
  }
  int collector(def o){ //parameter o is the input object 
    iterationSum = iterationSum + ((piData)o).iterations    //increment iterationSum
    withinSum = withinSum + ((piData)o).within  //increment withinSum by within
    return completedOK
  }

  int finalise(List p) {
    def pi = 4.0 * ((float) withinSum / (float) iterationSum)  // calculate pi
    println "Total Iterations: $iterationSum Points Within : $withinSum pi Value :$pi"
    return completedOK
  }
}
\end{lstlisting}
\end{minipage}

\subsection{Performance Analysis of Montecarlo $\pi$}
This analysis compares the performance of the versions given in Listings \ref{lst:listing3} and \ref{lst:listing4} and is shown in Table \ref{tab:table1} and using the constant values shown in Listing 1. The specification of the machines used in all analyses are given in Appendix \ref{sec:appendixC}, the crucial aspect is that the processor has 4 actual cores plus 4 hyper-threads.

The Speedup and Efficiency (percentage) values are calculated for the network in Listing \ref{lst:listing3} against the time for the Sequential version of the algorithm (Listing \ref{lst:listing4}).  It can be observed that, generally, there is a reduction in performance for the parallel version with just 1 worker process due to the overhead in setting up the parallel environment, but this is mostly no more than 2\% of the total execution time.  Typically, this percentage overhead gets smaller as the problem size increases.

\begin{figure}
        \centering
        \includegraphics[width=50mm]{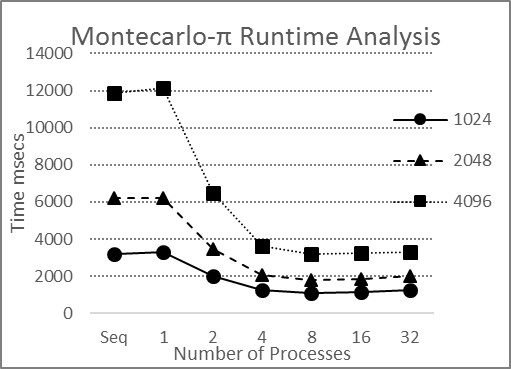}
        \caption{Montecarlo $\pi$ Runtime Analysis}
        \label{fig:graph1}
    \end{figure}

A simple count of the generated processes in Listing \ref{lst:listing3} is $workers + 4$.  Thus, the processor always has more actual processes than actual cores.  The additional four processes are mostly idle once all the $Workers$ are calculating.  Ideally, the Speedup should equal the number of cores, but this will never happen due to the additional processes.  It can also be observed that increasing the number of workers has little or no improvement once the number of worker processes equals the number of actual cores (see Figure \ref{fig:graph1}).  It can also be observed that as the amount of work increases by using more object instances, the Speedup and Efficiency do improve but this is limited to some extent by the size of the problem.  In the case of 8192 instances the application can make effective use of 4 processes after which improvement ceases.  We shall see that this is typical for all applications where the bigger the size of the problem in terms of data processed the easier it is to use more of the processes.

\begin{table}[]
\centering
    
\tiny

\begin{tabular}{|l|l|l|l|l|l|l|}
\hline
\textbf{Instances} & \multicolumn{2}{c|}{\textbf{1024}}     & \multicolumn{2}{c|}{\textbf{2048}}     & \multicolumn{2}{c|}{\textbf{4096}}     \\ \hline
\textbf{Processes} & \textbf{SpeedUp} & \textbf{Efficiency} & \textbf{SpeedUp} & \textbf{Efficiency} & \textbf{SpeedUp} & \textbf{Efficiency} \\ \hline
1                  & 0.98             &                     & 1.00             &                     & 0.98             &                     \\ \hline
2                  & 1.62             & 80.77               & 1.79             & 89.50               & 1.83             & 91.34               \\ \hline
4                  & 2.59             & 64.76               & 3.04             & 76.06               & 3.28             & 82.12               \\ \hline
8                  & 2.90             & 36.24               & 3.42             & 42.79               & 3.72             & 46.50               \\ \hline
16                 & 2.78             & 17.38               & 3.35             & 20.92               & 3.67             & 22.96               \\ \hline
32                 & 2.58             & 8.07                & 3.11             & 9.72                & 3.57             & 11.16               \\ \hline
\end{tabular}
\centering
\caption{  Performance of Montecarlo $\pi$ Object instances 1024, 2048 and 4096 
100,000 points created per object instance}
\label{tab:table1}

\end{table}

The effect of hyper-threading does not improve performance, an effect we shall see in most of the examples discussed in this paper.  In fact, the actual performance gets worse as the number of processes is increased beyond the number of threads and hyper-threads, as also observed in \cite{mskenney}. Thus, the user must be aware of the actual number of available cores and make sure that they are used up to that limit for the actual work to be undertaken.  It can also be observed that speedup and efficiency will differ for different sized problems and thus the user needs to be aware of this as well, but that is all part of becoming a confident parallel programmer.

\section{Groovy Parallel Patterns – Underlying Concepts}

The Groovy Parallel Patterns (GPP) Library \cite{kerridge_19a} contains three principal process types; $terminals$, $connectors$ and $functionals$.  The terminal processes provide a means of inserting data into a network of processes and finally removing results from a network.  Functional processes carry out one or more operations on the data object currently available to the process.  Connector processes provide a means of communicating data from one terminal or functional to another functional or terminal process.  Connector processes undertake no data processing, and  have two different forms, first, $spreaders$ have a single input and multiple output connections and $reducers$ have multiple input and one output connections. The names spreader and reducer are those suggested in \cite{danelutto_torquati_2013} and capture the role of the connector processes in terms of sending objects to subsequent parallel functionals and then collecting them into a single flow.   A $spreader$ enables data to be transferred from one source to many output destinations according to a distribution strategy appropriate to the problem solution.  In this manner, a data parallel architecture can be created by connecting a $spreader$ to a parallel set of $Worker$ processes, which is called a $group$.  Similarly, a set of $Worker$ processes can be connected one to the next without an intervening connector to create a task parallel $pipeline$ functional.  More complex functional architectures can then be created as either a pipeline of groups or a group of pipelines or other combinations of processes. A $reducer$ then takes inputs from several sources and outputs the input objects to a single destination process.  

All the processes in GPP use type declarations for all object properties and are compiled using the annotation @CompileStatic.  The extended Groovy type checking system means that the parser can ‘find’ types that match if there is some ambiguity, for example the Interface ChannelInput can refer to an internal or networked communication.  This has the effect of reducing the number of process variants.  However, the final system produced by the compiler, once gppBuilder has created the required code listings, is as efficient as that would have been produced by a Java compiler with all the data types fully specified.
Fundamental to the operation of the library is the concept of a \textit{DataClass} and its associated \textit{DataClassInterface}.  Every user defined object that is processed by a process network must \textit{extend DataClass}.  In addition, as shown in Listings \ref{lst:listing1}, \ref{lst:listing2} and \ref{lst:listing3}, processes are supplied with one or more $Details$ objects. Library processes use a variety of $Details$ objects that describe one or more data objects to the process.  These include, for example, the name of any methods that are used by the process together with any parameters passed to those methods.
As previously indicated, the GPP Library has components with behaviours which have been formally proved using CSPm \cite{gibson-robinson_armstrong_boulgakov_roscoe} and FDR4 \cite{gibson-robinson_armstrong_boulgakov_roscoe_2014}.  The description of the process behaviours is presented as each is discussed.  Finally, it is shown that combinations of the processes do have behaviours that can be automatically proved.  Any CSP based network can be formally shown to be correct using CSPm and FDR4, but this is often time consuming and can be challenging to undertake.  GPP removes the need for the user to check each network that is created because all the components have behaviours that follow the specification that is used to verify their behaviour.  Thus, provided the process network can be built using gppBuilder, the user can be assured the network will operate in a manner that is deadlock and livelock free.  A further advantage of the approach is that we can demonstrate that two different networks have the same behaviour.  None of the original skeleton libraries,  had this capability at the outset \cite{alg_skeleton} and only latterly are some groups working on algebras to help reasoning about the behaviour of their models.  The further advantage of FDR4 is that it is a fully automated tool.
\subsection{The DataClass Object}

The $DataClassInterface$ specifies the interface that all $DataClass$ objects must implement, it provides a set of constants that are required by developers, so user written code can interact with the processes through  which application data objects pass. As well as implementing the DataClassInterface, the DataClass object also defines an error message handler method that is called whenever an error is detected within and by the user code. This causes a message to be printed to the console with a user generated negative error code and the process network is then terminated.

\subsection{Details Objects}

Examples of Details classes include $DataDetails$ and $ResultDetails$, which are fundamental and examples of which are shown in Listing \ref{lst:listing1}.  Listing \ref{lst:listing7} gives the definition of the $DataDetails$ class and shows the default values for each property.

\begin{minipage}{\textwidth}
\begin{lstlisting}[caption = The DataDetails Class Definition,label=lst:listing7]
class DataDetails implements Serializable, Cloneable {
String dName = ""          // contains the name of the user defined data class
String dInitMethod = ""    // name of class initialise method
List dInitData = null      // parameter values used by the initialise method
String dCreateMethod = ""  // the name of the method that creates an object instance
List dCreateData = null    // parameter values used by the create method
String lName = ""          // the name of any local worker class the process utilises
String lInitMethod = ""    // the name of the local class' initialise method
List lInitData = null      // parameter values for the local class' initialise method
}
\end{lstlisting}
\end{minipage}

All user classes are constructed using an initialise method rather than the more usual public constructor.  Typically, the initialise method will be used to initialise constant, possibly $static$, properties of the class.  Parameters to methods are always passed in a $List$ structure so that the number of parameters can be varied both in number and type as required by the application.

\subsubsection{Result Details}
Listing 8 gives the definition of the $ResultDetails$ class and shows the default property values.  A Result Class also has its own initialise method and in addition two further methods.  The $collect$ method \{Listing \ref{lst:listing8}:5\} is passed a reference to an object from which it extracts the required data values and places them in the Result class object.  In this way, a sequence of objects can be processed.  The $finalise$ method produces \{Listing \ref{lst:listing8}:6\} the final output from the Result class object.

\begin{minipage}{\textwidth}
\begin{lstlisting}[caption = The ResultDetails Class Definition,label=lst:listing8]
class ResultDetails implements Serializable, Cloneable {
    String rName = ""           // the name of the class used to collate the results
    String rInitMethod = ""     // the name of the class initialise method
    List rInitData = null       // parameter values used by the initialise method      
    String rCollectMethod = ""  // the name of method used to collect the results
    String rFinaliseMethod = "" // the name of the method used to produce the result
    List rFinaliseData = null   // parameter values used by the finalise method  
    }
\end{lstlisting}
\end{minipage}

\subsection{Terminal Processes}
The terminal processes deal with the creation of data objects by an $Emit$ process and the collection of results by a $Collect$ process.

\subsubsection{The Emit Process Behaviour}
The properties of the Emit process are shown in Listing 9, where $emitDetails$ is a $DataDetails$ object that describes the data objects that are to be created which are written to the output channel.  The interface $CSProcess$ requires the creation of a $run()$ method which defines the behaviour of the process, but the user does not need to know the details of the method. 

\begin{minipage}{\textwidth}
\begin{lstlisting}[caption = The Properties of the Emit Process,label=lst:listing9]
class Emit implements CSProcess {    
    ChannelOutput output
    DataDetails emitDetails
}
\end{lstlisting}
\end{minipage}

Initially, the process creates an instance of the class defined by the $dName$ property of $emitDetails$ and then calls the object’s initialisation method $emitDetails.dInitMethod$ \{Listing \ref{lst:listing7}:3\}.  The main loop of the process creates a new instance of the emitted object and its associated $createMethod$  \{Listing \ref{lst:listing7}:5\} is called using $emitDetails.dCreateData$  \{Listing \ref{lst:listing5}:15-23\} as its parameters.  The return code indicates one of three possibilities; (1) $normalContinuation$, the object has been created and there are more objects to output, in which case, the object reference is written to the output channel; (2) $normalTermination$, there are no more objects to emit and the loop is terminated or (3) an error occurred, that will cause the whole system to exit with an error message.  After normal termination a $UniversalTerminator$ object is written to the $output$ channel to initiate network termination.  In Section 8 we shall discuss how this termination object can also be used to collate logging information.  

The process definition utilises the Groovy ‘.\&’ method pointer operator \cite{groovy_programming_language19}.  In the form used in this process and all other processes we can interpret: \\
\\
emitObject.\&”\$\{emitDetails.dCreateMethod\}”(emitDetails.dCreateData)\\
\\
as call the method with the name $emitDeatils.dCreateMethod$ defined in object $emitObject$ with the parameters $emitDetails.dCreateData$.  This mechanism has the benefit that all user defined methods can be invoked simply by using a String representation of the method’s name.  The $String$ representation does not have to be the same as the method’s name. 

\subsubsection{Specification of the Emit Process}

The definition of the Emit process is shown in CSPm Definition \ref{def:1}.  The datatype objects comprise a list of values A to E plus a terminating UT (Universal Terminator).  The $create()$ function specifies the internal operation of the $Emit$ and is equivalent to the $dCreateMethod$ referred to in \{ Listing \ref{lst:listing7}:5 \}.
\\

\begin{minipage}{\textwidth}
\begin{definition}{Datatype and Emit}
\begin{lstlisting}

datatype objects = A | B | C | D | E | A' | B' | C' | D' | E' | UT

create(A) = B      -- the specification of objects as they
create(B) = C      -- are created and emitted from the
create(C) = D      -- Emit process
create(D) = E 
create(E) = UT     -- the final terminating Universal Terminator

Emit(o) = a!o -> if o == UT then SKIP else Emit(create(o))
\end{lstlisting}
\label{def:1}
\end{definition}
\end{minipage}

$Emit(o)$ \{Definition \ref{def:1}:10\} defines the Emit process as output (!) an object $o$ from channel $a$ then test $o$ to determine its value.  If $o$ is the value $UT$ then the process adopts the behaviour $SKIP$, which implies that the process has terminated successfully.  Otherwise, the process adopts the behaviour of recursively calling Emit having evaluated $create(o)$.  It should be noted that $Emit$ will output the $UT$ value before doing $SKIP$ and terminating.

\subsubsection{The Collect Process Behaviour}
Listing \ref{lst:listing10} shows the properties of the Collect process, which has a local object defined by $resultDetails$, see Listing \ref{lst:listing2}, which is used to collect the results and process them.  Input objects are read from the channel $input$.

\begin{minipage}{\textwidth}
\begin{lstlisting}[caption = The Properties of the Collect Process,label=lst:listing10]
class Collect implements CSProcess {    
    ChannelInput input
    ResultDetails resultDetails
}
\end{lstlisting}
\end{minipage}
\\
The process creates and initialises an instance of the result object, after which an $inputObject$ is read.  The processing loop repeats until a $UniversalTerminator$ object is read.  The result object’s $collectMethod$ is called  \{Listing \ref{lst:listing6}:11-15\} with the $inputObject$ as a parameter and provided it $completedOK$ the next input object is read.  Once the loop has terminated the result object’s $finaliseMethod$  \{Listing \ref{lst:listing6}:17-21\} is called.  After which, the process terminates and thus the complete solution process network will now have terminated as all the preceding processes will also have terminated.  The $collectMethod$ extracts data from the input object and undertakes any necessary processing.  The $finaliseMethod$ undertakes any completion operation required and outputs the result as required by the solution.  

It should be noted that the class definition of the input object is not required by the process because it is only accessed in the result object’s $collect$ method, which is also user developed code.  In fact, most of the processes in the library do not require the class of any input object because all processing is undertaken by methods that have been written by the developer.

\subsubsection{Specification of the Collect Process}

The definition of the Collect process is given in CSPm Definition \ref{def:2}.  The process repeatedly inputs (?) a value from channel $d$ \{Definition  \ref{def:2}:1\} until the value $UT$ is read, when the process $Collect\_End$ is called.  $Collect\_End$ \{Definition  \ref{def:2}:2\} repeatedly outputs $true$ on the $finished$ channel.  FDR4 \cite{gibson-robinson_armstrong_boulgakov_roscoe} requires that it can make assertions about a set of processes.  If all the processes in a specification terminate, by behaving like $SKIP$, then the test for deadlock freedom cannot be undertaken.  Therefore, we create a specification that has a very simple terminal behaviour that is easy to test assertions against.
\\
\\
\begin{minipage}{\textwidth}
\begin{definition}{Collect Process}
\begin{lstlisting}
Collect() = d?o -> if o == UT then Collect_End() else Collect()
Collect_End() = finished!True -> Collect_End()

\end{lstlisting}
\label{def:2}
\end{definition}
\end{minipage}

\subsection{The Worker Process}
The simplest functional process is the $Worker$ process and its properties are shown in Listing \ref{lst:listing11} .  The channel $input$  \{Listing \ref{lst:listing11}:12\} is that from which objects to be processed are read and once processed the same object is written to the $output$ \{Listing \ref{lst:listing11}:13\} channel.  All objects are communicated by means of their object reference thereby removing the need for object copying unless explicitly required.  The GPP library processes have been built so that once a process has communicated an object reference it will never refer to that object again, thereby ensuring mutual exclusion of data access between processes.

\begin{minipage}{\textwidth}
\begin{lstlisting}[caption = The Properties of the Worker Process,label=lst:listing11]
class Worker implements CSProcess {    
    ChannelInput input            // channel from which object references are read
    ChannelOutput output          // channel to write references of updated object
    String function               // the name of the input object's method to be called
    List dataModifier = null      // any parameters required by the function method
    LocalDetails lDetails = null  // details of any local data class
    boolean outData = true        // when false only the finalised local class is output
    Barrier barrier = null        // used in groups to provide synchronisation
}
\end{lstlisting}
\end{minipage}

The main loop of the process reads an input object reference from channel $input$ and provided it is not an instance of the $UniversalTerminator$ object carries out the required $function$  \{Listing \ref{lst:listing11}:4\}.  The list of $dataModifer$ values, are used as parameters to the function.  The Worker process may have a local class used to hold intermediate results and this is described in the $LocalDetails$ \{Listing \ref{lst:listing11}:6\} object.  Assuming the function completes successfully then the reference of the updated object is written to the next process using the channel $output$.  Once the loop terminates the instance of $UniversalTerminator$ previously read is output thereby enabling the whole network to shut down in an orderly fashion.  In some cases, it may be required to output the local class rather than each input object, in which case the property $outData$  \{Listing \ref{lst:listing11}:7\} needs to be set $false$.  Finally, a group of Worker processes executing in a data parallel architecture, may need to ensure all workers in the group output their result only when all of them have completed the current calculation.  In this case the group will create a synchronisation barrier.  In this way, we can create an architecture like Valliant’s \cite{valiant_1990} bulk synchronous protocol BSP.  The group of $Worker$ processes will create the required synchronisation $barrier$ \{Listing \ref{lst:listing11}:8\} which is provided by the underlying JCSP library \cite{cspforjava}.

\subsubsection{Specification of the Worker Process}

The definition of the Worker processes is given in CSPm Definition \ref{def:3}. $Workers()$ comprises a parallel collection ($||$) of $N$ $Worker()$ processes \{Definition \ref{def:3}:9\}.  The alphabet or set of communication events each $Worker$ can observe is specified by $aW(x)$, thus $Worker(2)$ observes communications of the form $b.2$ and $c.2$, as specified in the $Worker(i)$ specification \{Definition \ref{def:3}:7-8\}.  $Worker(i)$ inputs (?) o from channel $b.i$; if $o$ is $UT$ then it is output on channel $c.i$ and the process then terminates by means of $SKIP$.  Otherwise, the value of $f(o)$ is output on $c.i$ and the process is called recursively.
\\
\\

\begin{minipage}{\textwidth}
\begin{definition}{Workers}
\begin{lstlisting}
N = 3
f(A) = A'       -- the function undertaken by Worker
f(B) = B'
f(C) = C'
f(D) = D'
f(E) = E'
Worker(i) = b?i.o -> if o == UT then (c!i.UT -> SKIP) else (c!i.f(o) -> Worker(i))
a_W(x) =  {|b.x, c.x|}  -- the alphabet used by each Worker(x)
Workers() =  || x : {0..N-1} @  [a_W(x)] Worker(x)
\end{lstlisting}
\label{def:3}
\end{definition}
\end{minipage}
\subsection{Connector Processes}

Connector processes are defined as either $spreaders$ or $reducers$.  The connector processes undertake no functional processing but simply connect terminal and functional processes together thereby forming a dataflow style architecture, similar to the components describe in RiscP2B \cite{danelutto_torquati_2013} and referred to in Chambers, Kerridge and Pedersen \cite{chalmers_jon_pederson_2016}.  The Connector processes terminate on receipt of a $UniversalTerminator$ and ensure  the required termination messages are sent to the processes connected to their output channels.

\subsubsection{Spreaders}

The processes are supplied in many different variants depending on the nature of the communication connections provided by the process as follows:
\begin{itemize}
\item $Any$ expects the any end of a channel to which many processes can read or write but only one object communication is processed at a time,
\item $List$ expects a channel list or array of channels accessed by index, either input or output,
\item $One$ expects a single connection to a process.
\end{itemize}

The nature of the process is defined by the central part of the name:
\begin{itemize}
    \item $Fan$ processes one object at a time and in the case of a List output will write the object to the next list out channel end in sequence,
\item $SeqCast$ outputs a single input value to all the outputs in sequence,
\item $ParCast$ outputs a single input value to all the outputs in parallel.
\end{itemize}

Thus, the spreader $OneFanAny$ reads an object from the single input channel and then outputs to just one of the many processes connected to the $any$ output channel. $OneFanList$, similarly reads an input object from its input channel and then outputs the object to the next element of the output channel array through which it iterates in a circular manner.  Writing to an $any$ channel means that the output object can be read by any of the idle processes waiting to read from the channel.  If all the reading processes are busy then the process waits, idle, until one of the reading processes is ready to communicate.  The descriptors $SeqCast$ and $ParCast$ cause the input object to be copied to all the output channels.  $SeqCast$ does the copy one output channel at a time in sequence.  $ParCast$ outputs the input object to all the output channels in parallel.  The spreaders that copy an input object to all the output channels use methods provided by groovyJCSP \cite{kerridge_18}.  They output a deep copy clone of the object that has been input.  The fact that objects are cloned, using a deep copy, is an important principle of the GPP Library because it means that within a single multi-core processor all objects are unique and thus it is safe to pass object references from one process to another.  If we could not guarantee that all objects are distinct then we could have more than one process accessing the same object reference at the same time and this would lead to problems with mutual exclusion between processes.  The library has been constructed to ensure that such problems do not occur, however, users must be aware that the default Java $clone()$ method is insufficient if the object being passed between processes contains further objects.  In this case the user must override the default clone method with one that undertakes a deep copy of the object.   The Groovy $@AutoClone$ annotation using the style $SERIALISATION$ provides a simple way of creating deep copy object clones  (\cite{Konig_2015}, page 260).
\\
\\
\begin{minipage}{\textwidth}
\begin{definition}{Generalised Spreader}
\begin{lstlisting}

Spread(i) = a?o -> if o == UT then 
                    ( b!i.UT -> Spread_End(i, (i+1) % N)) 
                   else 
                    ( b!i.o -> Spread((i+1) % N) )
Spread_End(s, n) = if s == n then SKIP else b!n.UT -> Spread_End(s, (n+1) % N )
\end{lstlisting}
\label{def:4}
\end{definition}
\end{minipage}

\subsubsection{Specification of a Spreader Process}

The general specification of a Spreader is shown in CSPm Definition \ref{def:4}. The process $Spread(i)$, inputs (?) an object $o$ from channel $a$, which is output (!) to channel $b.i$.  Spread is then called again recursively \{Definition \ref{def:4}:2-5\} accessing the next output channel in a round robin fashion until $UT$ is input from channel $a$.  The $UT$ object is output to the current $b.i$ channel after which the process $Spread\_End$ is invoked which causes $UT$ to be written to the remaining $b.i$ channels, which on completion causes the $Spread\_End$ process  \{Definition \ref{def:4}:6\} to behave as $SKIP$, which causes it to terminate.

\subsubsection{Reducers}

In all cases the reducer processes output to a single channel but input from many processes connected by either a one to many channel or a list (array) of channels.  Reading from a one to many channel means that as soon as one of the processes writing to the channel attempts to write then the reducer process will read that object unless it is already dealing with a previous input in which case the write request is queued in a FIFO structure and means that reads are processed in the order the writes occurred.
The $ALT$ construct, from groovyJCSP \cite{kerridge_18} provides a non-deterministic choice over the elements of an input channel list.  A $select$ method returns the index of the $inputList$ element that is chosen to communicate.  If no element is ready, then $select$ waits until one is ready and then processes the associated data.  If one element is ready, then that element is processed.  If more than one is ready, then the element is chosen according a number of selection criteria.  In the library we always chose a mechanism that allows equal bandwidth for all the channels, so called $fairSelect$.  It is also possible to input, either sequentially or in parallel from all the elements of a channel input list and output a single list that contains all the values that have been read.  Finally, reducers are provided that undertake merge operations so that objects can be input from the channel input list in a round robin fashion or in such a way as to ensure  the output objects are output in a sorted order assuming the data is presented on each input channel as a partial sorted data set.
It should be recalled that channel communication is synchronised, thus each $read$ or $write$ method call could implicitly create a delay as it waits for the corresponding opposite operation.  The library is constructed such that the connector processes do all the communication and thus provide a buffer between functional processes that consume most of the processing time.

\subsubsection{Specification of a Reducer Process}

The specification of a generalised Reducer process is given in CSPm Definition \ref{def:5}. A $Reducer()$ \{Definition \ref{def:5}:10\} comprises $N$ $Reduce()$ processes \{Definition \ref{def:5}:1\} formed as a replicated external choice ([]).  This allows an input on any of the c channels.  $Reduce(i)$ inputs an object on the i’th $c$ channel which is then output to channel $d$ until a $UT$ object is input.  Once UT has been input on one of the c channels the process $Reduce\_End()$ \{Definition \ref{def:5}:4-8\} is invoked, which iterates through the remaining $c$ channels outputting to the channel $d$ any further objects until $UT$ is input.


\begin{definition}{Generalised Reducer}
\begin{lstlisting}

Reduce(i) = c?i.o -> if o == UT then (Reduce_End(i, (i+1)%N) ) else ( d!o -> Reduce(i))

Reduce_End(s, n) = if s==n then d!UT -> SKIP else c?n.o -> 
                    if o == UT then 
                        Reduce_End(s, (n+1)%N )                         
                    else 
                        ( d!o -> Reduce_End(s, n) ) 

Reducer() = [] x: {0..N-1} @ Reduce(x)    -- replicated choice
\end{lstlisting}
\label{def:5}
\end{definition}

\subsection{Verification of the GPP Processing Model}
The remainder of the CSPm script that verifies the model is shown in CSPm Definition \ref{def:6}.

\begin{definition}{System Definition and Assertions}
\begin{lstlisting}

subtype emitObj = A | B | C | D | E | UT        --emitted objects
subtype fObj    = A' | B' | C' | D' | E' | UT   -- processed objects

channel a: emitObj            -- channels connecting the processes
channel b: {0..N-1}.emitObj   -- indicating the object types that
channel c: {0..N-1}.fObj      -- form the events on the channel
channel d: fObj          
channel finished : Bool       -- used during system refinement

a_A = {|a|} -- the alphabet (events) associated with each channel
a_B = {|b|} -- used in the construction of the whole system
a_C = {|c|}
a_D = {|d|}

System = (((Emit(A) [| a_A |] Spread(0)) [|a_B|] Workers()) 
            [| a_C |]  Reducer() ) [| a_D |] Collect()

TestSystem = finished!True -> TestSystem   --the model used for testing

assert (System \ {|a, b, c, d|}) [T= TestSystem   -- hide events on all channels but finished
assert (System \ {|a, b, c, d|}) [F= TestSystem
assert (System \ {|a, b, c, d|}) [FD= TestSystem
assert System :[deadlock free]
assert System :[divergence free]
assert System :[deterministic]
\end{lstlisting}
\label{def:6}
\end{definition}

Two subtypes are created, $emitObj$ and $fObj$, that subdivide the datatype objects created in CSPm Definition \ref{def:1}, to differentiate between objects that are emitted and those that have been processed by a $Worker$ process.  Each of the channels \{Definition \ref{def:6}:5-9\} used in the previous definitions are  defined together with the data they communicate.  Each channel has its own alphabet, or set of communication events \{Definition \ref{def:6}:11-14\}, defined, which can then be used in the parallel construction of the $System$.  The $System$ comprises the parallel composition \{Definition \ref{def:6}:16-17\} of the processes defined in CSPm Definitions \ref{def:1} to \ref{def:5}, identifying the alphabet of events that apply to each parallel.  The $TestSystem$ is then defined \{Definition \ref{def:6}:19\} which mimics the way in which the Collect process (CSPm Definition \ref{def:2}) finishes by repeatedly outputting the value $true$ to the $finished$ channel.  The $asserts$ \{Definition \ref{def:6}:21-26\} then check various aspects of the $System$, initially by showing that the $TestSystem$ is a refinement of the $System$, provided the order in which interactions on channels a to d are ignored.  The remaining $asserts$ check the $System$ for different capabilities, deadlock freedom, lack of divergence or livelock free and that its is deterministic.  All the assertions are satisfied, thereby giving confidence that the underlying design of the GPP library is correct, provided all the components are implemented in a manner that respects the above specifications.

This completes the description of the lowest level components of the library.  We could leave the library at this lowest level because we can construct any process network using the components described so far, however that would make using it very difficult and error prone.

\section{Higher Level Functionals}

Two basic functionals are provided, $groups$ and $pipelines$ together with $composites$ defining combinations of the other two.  A $matrix$ based set of process engines is also provided. 
\subsection{Groups}
Process networks within $groups$ provide the capability associated with a parallel-for statement by creating a parallel of $Worker$ processes.  They are typically used in data parallel applications where the same algorithm is applied to many instances of the same data.  The distinct types of group reflect the nature of the channel connections they have for input and output; any or channel lists.  Thus, options for: $AnyGroupAny, AnyGroupList, ListGroupList, ListGroupAny$ and a group $ListGroupCollect$ which contains a parallel of $Collect$ processes, are provided.  The number of parallel processes is specified by a property in the process’ constructor.  
\subsection{Pipeline}
The pipelines comprise two processes; $OnePipelineOne$ and $OnePipelineCollect$.  These processes are typically used in task parallel situations where a data object is passed through a series of processes (stages) each of which can be executed in parallel.  The only difference is that in $OnePipelineCollect$ the last stage of the pipeline is a $Collect$ process.  Pipelines always process a single input channel and a single output channel and must always have at least two stages.  All the internal communication channels are created automatically.
\subsection{Composite}
The $composite$ processes define process networks that are either a pipeline of groups or a group of pipelines.  These process networks are characterized by the number of workers in each group and the number of pipeline stages.  Any group stage that has a List type output can also create a synchronising barrier that ensures all processes in the parallel complete the process loop before any of them can write their output to the next stage.
\subsection{Matrix}
The GPP library contains two specific matrix-based architectures, both of which assume that the data in the matrix is partitioned into distinct subsets which can be processed independently.  Thus, each partition can be processed by its own core before being recombined with the other parts to form a complete solution.  The processing engine comprises a root node together with many worker nodes, the number of which is specified by the user.  Internally these engines access the data in a shared manner so that data is not copied but the user has no direct access to the shared data; they simply specify how the data should be partitioned over the worker nodes.  One of the engines assumes that the matrix will be processed by a series of iterations, each of which updates the initial matrix.  The other is based on image kernel processing and assumes that each matrix is processed by one engine and then passed onto another engine for another operation.

\section{Further Multi-Core Examples}

This section provides further examples of the use of the GPP library using some commonly encountered problems and ones that utilise a variety of process networks and other processes contained in the library.  Some of these problems could benefit from the use of a cluster of multi-core workstations but this aspect is discussed in Section 7 where a solution to the Mandelbrot Set is presented.  All the examples are taken from the GPP library demonstrations, which contain further versions using different networks and other examples \cite{kerridge_19}.

\subsection{Concordance: Basic Map – Reduce Algorithm}

The production of a concordance \cite{bible_concordance} requires the processing of a large text to find the location of every occurrence of a specific word or string of words.  For each string of words up to some maximum size $N$, a file should be produced giving the word strings and the locations where that string is found in the text.  An output should only be produced provided the number of locations is greater than or equal to the value of $minSeqLen$ and the approach adopted is:
\begin{enumerate}
    \item Read in the text file, remove extraneous punctuation from each word and then calculate an integer value corresponding to the sum of the letter codes in the word.  The sum value and the word are stored in $static$ data structures in an object.  Instances of this object are created for each string length value from 1 to $N$.  This function is carried out by an $init$ method that initialises $static$ variables and a create method used to create each instance corresponding to the individual values on $n$ in $1 .. N$.  
    
    \item For each value $n$, create another data structure containing the sum of $n$ integer values starting with the first word value then the second and so on for the whole text.  This creates a $valueList$ of integer values that represents the sum of $n$ words for each location in the file. 
    \item Process the $valueList$ to find equal values to determine the locations of the value in the file.  This is stored in a map called $indicesMap$.
    \item The $indicesMap$ can now be processed to determine the word string associated with each value by referencing the word data structure saved in the object.  In some cases, the same value will refer to different strings of words and so this must be disambiguated.  The resulting map called $wordsMap$ contains a string of words together with the locations where it was found.  
    \item Provided the number of occurrences is greater than $minSeqLen$ then this can be output to a file, one for each value of $n$.

\end{enumerate}

\begin{figure}
        \centering
        \includegraphics[width=100mm]{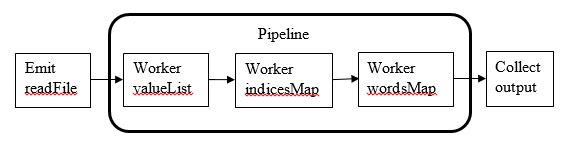}
        \caption{Pipeline Structure for Concordance}
        \label{fig:6-1}
    \end{figure}

The phases 2 to 4 can be carried out in a pipeline, see Figure \ref{fig:6-1}, with each stage of the pipeline undertaking a different operation on the object for its value of $n$.  The first phase can be undertaken in an $Emit$ process.  Phase 5 can be undertaken in a $Collect$ process or as part of the pipeline.  The functions to be carried out in each phase are the names already given; $valueList$, $indicesMap$, $wordsMap$. 

Each of the phases 2 to 4 comprises a map-reduce style \cite{mapreduce} process.  It is organised as a sequence of operations one following the next.  Thus, in this relatively simple manner GPP can support Map-Reduce style processing.

The algorithm was tested on a file containing the Bible \cite{bible} , which contains 802,000 words and is 4.6 mbytes in size.  The output for $N$ = 6 totals 26 mbytes.  Thus, the application is I/O bound but we can use it to explore different parallel networks and thereby compare their relative performance and against a sequential solution.  All the solutions use the same Detail object definitions as shown in Listing \ref{lst:listing12}.

\begin{minipage}{\textwidth}
\begin{lstlisting}[caption = Detail Object Definitions,label=lst:listing12]

def fileName = "src\\demos\\concordance\\bible.txt"        //the name of the source text file
def outFileName = "src\\demos\\concordance\\bibleOutput"   // the output file name
int N = 6                     // maximum number of words in a string
int minSeqLen = 2             // minimum number of occurrences

def dDetails = new DataDetails( dName: concordanceData.getName(),
                                dInitMethod: concordanceData.init,
                                dInitData: [N, fileName, outFileName],
                                dCreateMethod: concordanceData.create )
def rDetails = new ResultDetails( rName: concordanceResults.getName(),
                                  rInitMethod: concordanceResults.init,
                                  rInitData: [minSeqLen],
                                  rCollectMethod: concordanceResults.collector,
                                  rFinaliseMethod: concordanceResults.finalise )
\end{lstlisting}
\end{minipage}

Listing \ref{lst:listing13} shows the use of a composite process called $GroupOfPipelineCollects$.  This creates a parallel set of pipelines each comprising a 3-stage pipeline with a $Collect$ process as a further final stage. There are $groups$ parallel pipelines.  The object $resultDetails$ contains a copy of the $rDetails$ object for each instance of the pipeline, so that each $Collect$ process has access to the results collection methods.

\begin{minipage}{\textwidth}
\begin{lstlisting}[caption = Formulation as a Group of Pipeline Collects from functionals.composites,label=lst:listing13]
int groups = 2
List <ResultDetails>  resultDetails = []
for ( g in 0..< groups) resultDetails << rDetails

def emitter = new Emit( eDetails: dDetails)
def fanOut = new OneFanAny(destinations: groups)
def GoPconcordance = new GroupOfPipelineCollects( stages: 3,
                                                  rDetails: resultDetails,
                                                  stageOp: [cd.valueList, 
                                                            cd.indicesMap, 
                                                            cd.wordsMap],
                                                  groups: groups)
\end{lstlisting}
\end{minipage}

Listing \ref{lst:listing14} shows the use of a supplied pattern, $TaskParallelOfGroupCollects$, which creates a pipeline of $workers$ groups of $Worker$ processes.  Each group carries out the same operation but on a different data object.  The pipeline is $stages$ long with a $Collect$ process as an additional final stage.

\begin{minipage}{\textwidth}
\begin{lstlisting}[caption = Formulation using the pattern TaskParallelGroupOfCollects,label=lst:listing14]
int workers = 2
List <ResultDetails>  resultDetails = []
for ( g in 0..< workers) resultDetails << rDetails

def PoGconcordance = new TaskParallelOfGroupCollects(eDetails: dDetails,
                                                     rDetails: resultDetails,
                                                     stages: 3,
                                                     stageOp: [cd.valueList, 
                                                               cd.indicesMap, 
                                                               cd.wordsMap],
                                                     workers: workers )
                                                     
\end{lstlisting}
\end{minipage}

The total code length of the Concordance system comprises about 35 lines of GPP Library coding.  The same application using JCSP and groovyJCSP coding requires around 145 lines of coding  \cite[chapter~24]{kerridge_2014}. In both cases the sequential coding is ignored and it is identical for both systems.  The GPP coding is about one-quarter the length of the same system using normal coding practices.  Further, the GPP version has an underlying behaviour that has a formally proved structure.  The JCSP/groovyJCSP version would require the user to undertake this analysis themselves.

\subsubsection{Equivalence of Process Structures}

Using the CSPm Definitions \ref{def:1},\ref{def:2}, \ref{def:4} and \ref{def:5} we can create two versions of the structures as shown in CSPm Definition \ref{def:7} .  The alphabet of any channel z is indicated by $a\_Z$ \{Definition \ref{def:7}:5,14-16\}.  Each of the Workern processes has their own function $f(o)$ but are otherwise identical to that shown in CSPm Definition \ref{def:3}.  The systems being refined are a pipeline of three worker groups (PoG), each with two workers \ref{def:7}:22-24) and a group of two parallel pipelines each comprising three worker stages (GoP) \{Definition \ref{def:7}:9-12\}. The assertions \{Definition \ref{def:7}:25-27\} test the equivalence of the two systems, which do have the same behaviour, provided the internal ordering of internal events is hidden (using the $\backslash$ notation).
\\
\\
\begin{minipage}{\textwidth}
\begin{definition}{Concordance System Refinement}
\begin{lstlisting}
Worker1(i) = b?i.o -> if o == UT then ( c!i.UT -> SKIP ) else ( c!i.f1(o) -> Worker1(i) )
Worker2(i) = c?i.o -> if o == UT then ( d!i.UT -> SKIP ) else ( d!i.f2(o) -> Worker2(i) )
Worker3(i) = d?i.o -> if o == UT then ( e!i.UT -> SKIP ) else ( e!i.f3(o) -> Worker3(i) )

a_Pipe(x) = {| b.x, c.x, d.x, e.x |}   -- the alphabet for a Pipeline of Workers(x)

Pipe(i) = (Worker1(i) [| a_C |] Worker2(i)) [| a_D |] Worker3(i)  -- pipe definition

GoP() = || x: {0..1} @ [ a_Pipe(x) ] Pipe(x)    -- define two pipes {0..1}

GoPSystem = ( ( ( Emit(A) [| a_A |] Spread(0) )  
                [| a_B |]  GoP() ) [| a_E |] Reducer() ) [| a_F |] Collect()

a_G1(x) = {| b.x, c.x |} -- the alphabet for each of the Groups
a_G2(x) = {| c.x, d.x |}
a_G3(x) = {| d.x, e.x |}

Group1() = || x:{0..1} @ [ a_G1(x) ] Worker1(x) -- groups comprising two parallel workers
Group2() = || x:{0..1} @ [ a_G2(x) ] Worker2(x)
Group3() = || x:{0..1} @ [ a_G3(x) ] Worker3(x)

PoG() = (Group1() [| a_C|] Group2()) [|a_D|] Group3()
PoGSystem = ( ( ( Emit(A) [| a_A |] Spread(0) ) 
    [| a_B |]  PoG() ) [| a_E |] Reducer() ) [| a_F |] Collect()
assert (PoGSystem \ {|a, b, c, d, e, f|}) [T= (GoPSystem \ {|a, b, c, d, e, f|})
assert (PoGSystem \ {|a, b, c, d, e, f|}) [F= (GoPSystem \ {|a, b, c, d, e, f|})
assert (PoGSystem \ {|a, b, c, d, e, f|}) [FD= (GoPSystem \ {|a, b, c, d, e, f|})
\end{lstlisting}
\label{def:7}
\end{definition}
\end{minipage}

\subsubsection{Concordance Performance Analysis}

\begin{figure}
        \centering
        \includegraphics[width=80mm]{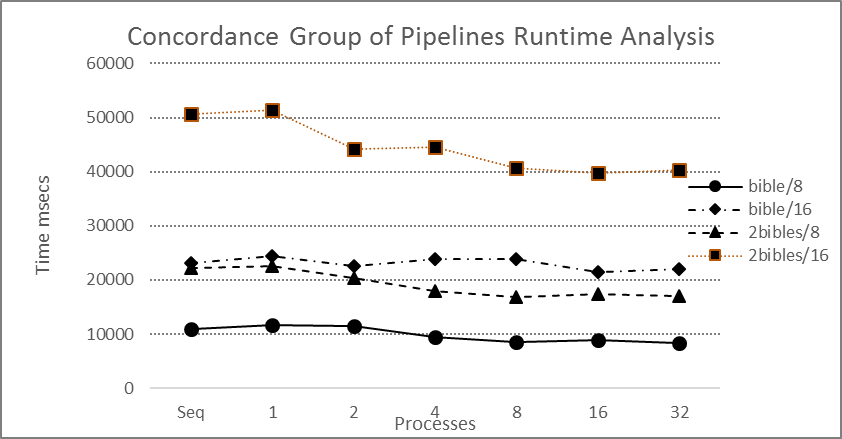}
        \caption{Runtime Analysis for Concordance Group of Pipelines Version
Texts: bible and 2bibles for Word String Lengths of N = 8 and 16 
Sequential together with 1,2,4,8,16 and 32 parallel groups}
        \label{fig:graph2}
    \end{figure}

The analysis of the performance of the different solutions is shown in Table \ref{tab:table2} and \ref{tab:table3}. The networks were assessed using N = 8 and 16 for two different texts, the Bible \cite{bible} and two concatenated copies of the Bible.  It can be observed that GoP version (Table 2) has a slightly better overall performance but neither shows a great performance improvement over the sequential solution, because the problem is I/O bound.  The somewhat erratic performance is shown more clearly in Figure \ref{fig:graph2}.  The PoG version has a similar performance.   As can be seen the performance improves with the larger 2bibles text and the larger value of N =16. Figure \ref{fig:graph4} shows the effect of adding extra parallel pipelines to the architecture shown in Figure \ref{fig:graph5}, which shows just 1 pipeline.  Thus 2 parallel pipelines will invoke a total of 8 processes for the pipelines plus a further two for the Emit and Collect processes.  Given the test machine has 4 cores and 4 hyper-threads subsequent increases in the number of parallel pipelines is likely to produce little or no performance improvement.

In Section 8.1, an analysis of logging output will be used to determine how parts of the network can be improved.

\begin{table}[]
\tiny
\centering
\begin{tabular}{|l|l|l|l|l|l|l|l|l|}
\hline
\textbf{GoP}       & \textbf{Text/N}  & \textbf{bible/8}    & \textbf{Text/N}  & \textbf{bible/16}   & \textbf{Text/N}  & \textbf{2bibles/8}  & \textbf{Text/N}  & \textbf{2bibles/16} \\ \hline
\textbf{Processes} & \textbf{SpeedUp} & \textbf{Efficiency} & \textbf{SpeedUp} & \textbf{Efficiency} & \textbf{SpeedUp} & \textbf{Efficiency} & \textbf{SpeedUp} & \textbf{Efficiency} \\ \hline
1                  & 0.94             &                     & 0.95             &                     & 0.98             &                     & 0.98             &                     \\ \hline
2                  & 0.96             & 47.83               & 1.02             & 50.90               & 1.10             & 54.89               & 1.15             & 57.37               \\ \hline
4                  & 1.16             & 28.91               & 0.96             & 24.08               & 1.24             & 31.10               & 1.13             & 28.37               \\ \hline
8                  & 1.27             & 15.92               & 0.97             & 12.13               & 1.32             & 16.50               & 1.25             & 15.56               \\ \hline
16                 & 1.22             & 7.65                & 1.07             & 6.70                & 1.28             & 8.02                & 1.28             & 7.98                \\ \hline
32                 & 1.31             & 4.08                & 1.04             & 3.26                & 1.31             & 4.10                & 1.25             & 3.92                \\ \hline
\end{tabular}
\caption{Performance Analysis for Group of Pipeline (GoP) Solution to the Concordance Problem Compared to Sequential Invocation for two texts, bible and 2bibles Using Word String Lengths of $N$ = 8 and 16
}
\label{tab:table2}
\end{table}

\begin{table}[]
\tiny
\centering
\begin{tabular}{|l|l|l|l|l|l|l|l|l|}
\hline
\textbf{PoG}       & \textbf{Text/N}  & \textbf{bible/8}    & \textbf{Text/N}  & \textbf{bible/16}   & \textbf{Text/N}  & \textbf{2bibles/8}  & \textbf{Text/N}  & \textbf{2bibles/16} \\ \hline
\textbf{Processes} & \textbf{SpeedUp} & \textbf{Efficiency} & \textbf{SpeedUp} & \textbf{Efficiency} & \textbf{SpeedUp} & \textbf{Efficiency} & \textbf{SpeedUp} & \textbf{Efficiency} \\ \hline
1                  & 0.90             &                     & 0.93             &                     & 0.95             &                     & 0.92             &                     \\ \hline
2                  & 0.96             & 47.97               & 1.00             & 50.07               & 1.06             & 53.15               & 1.07             & 53.38               \\ \hline
4                  & 1.11             & 27.85               & 0.90             & 22.51               & 1.19             & 29.74               & 1.11             & 27.68               \\ \hline
8                  & 1.18             & 14.76               & 0.91             & 11.39               & 1.28             & 16.02               & 1.16             & 14.44               \\ \hline
16                 & 1.23             & 7.70                & 1.03             & 6.44                & 1.29             & 8.09                & 1.24             & 7.73                \\ \hline
32                 & 1.22             & 3.80                & 1.05             & 3.27                & 1.28             & 4.01                & 1.27             & 3.98                \\ \hline
\end{tabular}
\caption{Performance Analysis for Pipeline of Group (PoG) Solution to the Concordance Problem }
\label{tab:table3}
\end{table}

\subsection{Jacobi’s Method: Dense Linear Algebra}

Jacobi’s method \cite{jacobi_algorithm} is a means of solving systems of diagonally dominant simultaneous equations.  It is easier to parallelise than the method due to Gauss \cite{gauss_algorithm}, which is refinement of Jacobi’s method, but which updates in place rather than Jacobi’s method which does an update at the end of each iteration.  The method starts with an initial guess for each variable and then successively refines the guess either for a specific number of iterations or until an error margin is reached.  Data for testing the algorithm was created randomly but because the solution was known it is possible to check the algorithm works correctly.  A set of test files was created with dimension 1024, 2048, 4096 and 8192 variables / equations.  The test files are guaranteed to be diagonally dominant,  a requirement of Jacobi's method.

Listing \ref{lst:listing15} gives the definition of the network required to solve the problem.  It is assumed that the data is read in from a text file ($fileName$  \{Listing \ref{lst:listing15}:1\} and copied into an object of type $jacobiData$ using the object’s $initMethod$ and $createMethod$ as specified in $eDetails$  \{Listing \ref{lst:listing15}:4-7\}.  The text file also contains the known solution to the problem and this is also stored in the object.  The input text file may contain more than one set of equations of different dimensions and these will be passed to the $MultiCoreEngine$ process as a stream of data.  The object $jacobiResults$ \{Listing \ref{lst:listing15}:8-11\} stores the resulting variable values and does a check to ensure correctness in the collector method.  Obviously, in a real situation the ability to check for correctness is not available.

\begin{minipage}{\textwidth}
\begin{lstlisting}[caption = Jacobi Method Network,label=lst:listing15]
def fileName = "src\\demos\\jacobi\\jacobi.txt"
double margin = 1.0E-16

def eDetails = new DataDetails (dName: jacobiData.getName(),
                                dCreateMethod: jacobiData.createMethod,
                                dInitMethod: jacobiData.initMethod,
                                dInitData: [fileName])
def rDetails = new ResultDetails(rName: jacobiResults.getName(),
                                 rInitMethod: jacobiResults.init,
                                 rCollectMethod: jacobiResults.collector,
                                 rFinaliseMethod: jacobiResults.finalise)

def emit = new Emit( eDetails: eDetails)
def mcEngine = new MultiCoreEngine (nodes: nodes,
                                    errorMargin: margin,
                                    finalOut: true,
                                    partitionMethod: jacobiData.partitionMethod,
                                    calculationMethod: jacobiData.calculationMethod,
                                    errorMethod: jacobiData.errorMethod,
                                    updateMethod: jacobiData.updateMethod )
def collector = new Collect(rDetails: rDetails)
\end{lstlisting}
\end{minipage}

The $MultiCoreEngine$ \{Listing \ref{lst:listing15}:14-20\} process comprises a Root node and as many worker Nodes, specified by $nodes$.  Typically, the number of Nodes will be the number of available cores in the processor.  In this example, the network iterates until the error margin specified is achieved for all variable values by taking the absolute value of the difference between the latest iteration and the previous one.  The user must specify the partitioning of the input data such that the index of each node specifies the partition it is to operate upon.  The $calculationMethod$ \{Listing \ref{lst:listing15}:18\} specifies the operation to be undertaken in this case based upon the description \cite{jacobi_algorithm}.  The calculation is carried out in the nodes, such that each node only undertakes the operation for the values in its partition but can access all the other current guesses as required.    There is only one copy of the data which is shared between the Root and Node processes in such a way that the Nodes only write data associated with their partition but can read all the other required data.  This internal structuring is hidden from the user apart  from having to specify how the data is to be partitioned, by $partitionMethod$  \{Listing \ref{lst:listing15}:17\}.

Once all the nodes have completed their calculations, the Root node resumes to undertake the following sequence.  The $errorMethod$ \{Listing \ref{lst:listing15}:19\} determines whether each new guess is within $errorMargin$ of the previous one and if another iteration is required returns the value true, false otherwise.  The $updateMethod$  \{Listing \ref{lst:listing15}:20\} is used to transfer the latest guess from its location into the place for the last guess to be able to start another iteration.

\begin{figure}
        \centering
        \includegraphics[width=80mm]{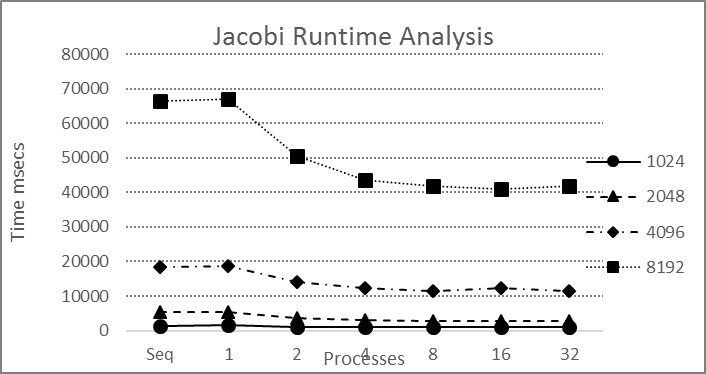}
        \caption{Jacobi Runtime Analysis}
        \label{fig:graph3}
    \end{figure}

\begin{table}[]
\tiny
\centering
\begin{tabular}{|l|l|l|l|l|l|l|l|}
\hline
\textbf{Jacobi}    & \textbf{1024}       & \textbf{Equations} & \textbf{2048}       & \textbf{Equations} & \textbf{4096}       & \textbf{Equations} & \textbf{8192}       \\ \hline
\textbf{Nodes} & \textbf{Efficiency} & \textbf{SpeedUp}   & \textbf{Efficiency} & \textbf{SpeedUp}   & \textbf{Efficiency} & \textbf{SpeedUp}   & \textbf{Efficiency} \\ \hline
1                  &                     & 0.98               &                     & 0.98               &                     & 0.99               &                     \\ \hline
2                  & 58.87               & 1.48               & 74.07               & 1.30               & 64.86               & 1.32               & 65.83               \\ \hline
4                  & 32.62               & 1.81               & 45.28               & 1.50               & 37.50               & 1.53               & 38.23               \\ \hline
8                  & 17.26               & 2.06               & 25.70               & 1.60               & 20.05               & 1.59               & 19.92               \\ \hline
16                 & 8.76                & 1.94               & 12.09               & 1.50               & 9.39                & 1.62               & 10.12               \\ \hline
32                 & 4.66                & 1.98               & 6.19                & 1.59               & 4.97                & 1.59               & 4.96                \\ \hline
\end{tabular}
\caption{Performance of Jacobi Network For 1024, 2048, 4096 and 8192 sets of equations Coefficients created randomly and guaranteed diagonal dominance}
\label{tab:table4}
\end{table}

Table \ref{tab:table4} provides an analysis of the performance against a sequential invocation of the same methods.  In all cases the Emit, Collector and Root node has been discounted from the number of processes, which simply records the number of worker Nodes.  Figure \ref{fig:graph3} shows the runtime analysis of the system for various sizes of data sets.
The performance of the system is reasonable given that there is a sequential phase where the errors values are determined, and new values are moved within the data, prior to the next iteration.

The performance improvement is more noticeable for the larger datasets.  More importantly the time increase for 8192 over 4096 equation is 71\% of the longer time and the increase in the number of coefficients is 75\%, showing that the increase is close to linear.  The test machine had 4 cores and thus further improvement after that is minimal.

\subsection{Planetary Movement: N-body Problem}

The N-body problem solution shown in Listing \ref{lst:listing16}, also utilises the $MultiCoreEngine process$, used in the solution to the Jacobi system.  In this instance though, the algorithm just runs for a fixed number of iterations, as the concept of an error margin is not appropriate.  Two different implementations were used, one using an array of body objects and the other using a matrix to hold all the data values.  The difference in performance was minute, so only one set of results are presented.    The algorithm is based upon that described in \cite{nbody_algorithm}, where the Java code is copied directly into the data objects used in the solution.  The file indicated by $readPath$  \{Listing \ref{lst:listing16}:5\} contains sufficient data for 10,000 planets.  The different sized problems simply take the required number of data points from the file.  The final state of a running of the system is output to another file, $writePath$ \{Listing \ref{lst:listing16}:6\}.  Each running of the system has the output compared with a sequential execution of the problem to check that all the solutions are identical, using an external program.

\begin{minipage}{\textwidth}
\begin{lstlisting}[caption = Planetary Movement Network,label=lst:listing16]
int nodes = 2			// number of worker nodes used in the MultiCoreEngine
int N = 100			// the number of bodies
int iterations = 100		// the number of iterations
double dt = 1e11		// the interval delta between iterations
String readPath = "src/demos/nbody/planets_list.txt"
String writePath = "src/demos/nbody/result_${iterations}_${N}_planets_${nodes}.txt"

def eDetails = new DataDetails (dName: nBodyData.getName(),
                                dCreateMethod: nBodyData.createMethod,
                                dInitMethod: nBodyData.initMethod,
                                dInitData: [readPath, N, dt])
def rDetails = new ResultDetails(rName: nBodyResult.getName(),
                                 rInitMethod: nBodyResult.init,
                                 rInitData: [writePath],
                                 rCollectMethod: nBodyResult.collector,
                                 rFinaliseMethod: nBodyResult.finalise)

def emit = new Emit( eDetails: eDetails)
def mcEngine = new MultiCoreEngine (nodes: nodes,
                                    finalOut: true,
                                    iterations: iterations,
                                    partitionMethod: nBodyData.partitionMethod,
                                    calculationMethod: nBodyData.calculationMethod,
                                    updateMethod: nBodyData.updateMethod )
def collector = new Collect(rDetails: rDetails)
\end{lstlisting}
\end{minipage}

The role of $nBodyData$ and $nBodyResults$ \{Listing \ref{lst:listing16}:8-11,12-16\} is the same as in the Jacobi system with the object methods used in the same way.  Note the parameter values passed into the system by means of $dInitData$ (Listing \{\ref{lst:listing16}:11\}) and $rInitData$ \{Listing \ref{lst:listing16}:4\} properties.  The data has to be partitioned in some user defined manner and all nodes have access to all the data but can only write data in their own partition.  Table \ref{tab:table5} shows the performance of the array of objects version of the problem.  Once again we observer that the performance improvement up to the  number of cores is as expected and thereafter drops off, especially in terms of efficiency.

\begin{table}[h]
\tiny
\centering
\begin{tabular}{|l|l|l|l|l|l|l|}
\hline
\textbf{N-body}    & \textbf{Bodies}  & \textbf{2048}       & \textbf{Bodies}  & \textbf{4096}       & \textbf{Bodies}  & \textbf{8192}       \\ \hline
\textbf{Nodes} & \textbf{SpeedUp} & \textbf{Efficiency} & \textbf{SpeedUp} & \textbf{Efficiency} & \textbf{SpeedUp} & \textbf{Efficiency} \\ \hline
1                  & 0.94             &                     & 1.07             &                     & 1.02             &                     \\ \hline
2                  & 1.56             & 89.50               & 2.09             & 104.66              & 1.97             & 98.43               \\ \hline
3                  & 1.84             & 80.78               & 2.80             & 93.36               & 2.74             & 91.20               \\ \hline
4                  & 1.89             & 62.55               & 3.29             & 82.20               & 3.30             & 82.53               \\ \hline
8                  & 2.16             & 38.15               & 3.97             & 49.60               & 3.87             & 48.32               \\ \hline
16                 & 2.17             & 16.14               & 3.72             & 23.25               & 3.71             & 23.18               \\ \hline
32                 & 1.90             & 7.60                & 3.00             & 9.39                & 3.78             & 11.83               \\ \hline
\end{tabular}
\caption{ Analysis of the Nbody Problem For 2048, 4096 and 8192 bodies Taken from a file of 10,000 randomly generated bodies
}
\label{tab:table5}
\end{table}

\begin{figure}
        \centering
        \includegraphics[width=80mm]{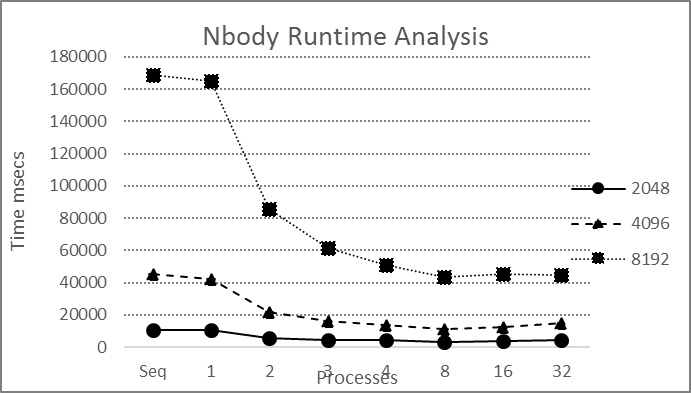}
        \caption{Timing Analysis for the Nbody Problem}
        \label{fig:graph4}
    \end{figure}
    
Figure \ref{fig:graph4} shows that runtime analysis where the improvement in performance as additional nodes are added is more obvious.  The effect of the sequential update phase is much less in this case, compared with the Jacobi solution as there is no evaluation of errors. The drop off in speedup / efficiency is pronounced as the number of nodes goes from 4 to 8 because the processor only has 4 cores.

\subsection{Image Processing: Kernel Based Stencil Algorithms}

Stencil algorithms are commonly use in image processing examples \cite{image_filtering}.  The required processing is very similar to the $MultiCoreEngine$ except that images are often put through a sequence of operations and there is also a need to double buffer the data objects.  Thus, assuming a stream of input images, we need to create a sequence of processing stages.  These are the additional capabilities provided by the $ImageEngine$ process. The network shown in Listing \ref{lst:listing17}  shows a sequence of two $ImageEngine$ processes, the first, $engine1$  \{Listing \ref{lst:listing17}:29-31\} coverts a coloured image into greyscale.  Then $engine2$ undertakes and edge detection operation before the image is output to the collector.  Two edge detection kernels are available one 3x3 and the other 5x5. 

The image object contains the definition of all the user defined operations.  The $initMethod$ initialises any properties of the object as required.  The $createMethod$ is used to read in an image file and store it in the desired format for processing and is passed the values of the $inFile$ and $outFile$  \{Listing \ref{lst:listing17}:22\} parameters.  The $partitionMethod$ \{Listing \ref{lst:listing17}:30\} splits the data so that individual blocks of the image can be processed in parallel.  This method is only called once in the first engine to process the image.  The $greyScaleMethod$ \{Listing \ref{lst:listing17}:31\} is used to transform a coloured image into a grey scale.  The $convolutionMethod$ undertakes the stencil operation using a kernel matrix passed as $convolutionData$  \{Listing \ref{lst:listing17}:33-34\}.  The actual algorithm used to undertake the operation is defined by the user.  Finally, the method $updateImageIndex$  \{Listing \ref{lst:listing17}:35\} is used to indicate which of the two buffers should be used for the next operation if any, it is only specified when the operation undertaken uses the double buffer capability.

\begin{figure}[h]
 
        \centering
        \includegraphics[width=70mm]{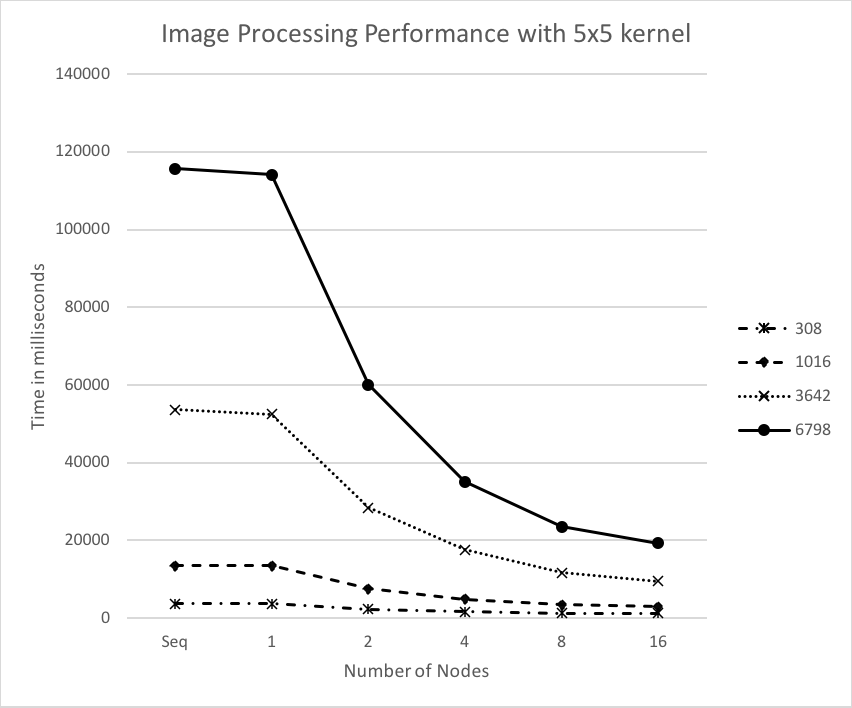}
        \caption{Runtime Performance of Image Processing with 5x5 Edge Detection Kernel
Image sizes of 308, 1016, 3642 and 6798 kb are all versions of the same image
}
        \label{fig:graph5}
    \end{figure}

For the analysis, an initial 24 megapixel image (6000x4000) containing 6798 KB, was processed to create a series of smaller sized images with the X dimension reduced from 6000 to 1024, 2048 and 4096 pixels.  The resulting file sizes were 308, 1016, 3642 KB respectively.  It can be seen from Figure \ref{fig:graph5} and Table \ref{tab:table6} that the performance improvement up to 8 nodes is reasonable, especially for larger image sizes.  Smaller image files were also produced with 256 and 512 pixels, but these had worse performance for the parallel versions over the sequential version.  The difference between the run time for the 3x3 versus the 5x5 kernel was increase of between 8\% and 20\% longer than the 3x3 version.  Given there is an increase of 1.56 in the number of calculation undertaken per pixel shows that performance hit using the 5x5 kernel is not that large.

\begin{table}[]
\tiny
\begin{tabular}{|c|c|c|c|c|c|c|c|c|}
\hline
\multirow{2}{*}{\begin{tabular}[c]{@{}c@{}}SizeKB\\
 Nodes\end{tabular}} & \multicolumn{2}{c|}{308} & \multicolumn{2}{c|}{1016} & \multicolumn{2}{c|}{3642} & \multicolumn{2}{c|}{6798} \\ \cline{2-9} 
                                                                              & Speedup   & Efficiency   & Speedup    & Efficiency   & Speedup    & Efficiency   & Speedup    & Efficiency   \\ \hline
1                                                                             & 1.01      & 1.01         & 1.00       & 1.00         & 1.02       & 1.02         & 1.01       & 1.01         \\ \hline
2                                                                             & 1.59      & 0.79         & 1.78       & 0.89         & 1.89       & 0.94         & 1.93       & 0.96         \\ \hline
4                                                                             & 2.30      & 0.57         & 2.79       & 0.70         & 3.05       & 0.76         & 3.30       & 0.82         \\ \hline
8                                                                             & 3.01      & 0.38         & 3.88       & 0.48         & 4.58       & 0.57         & 4.91       & 0.61         \\ \hline
16                                                                            & 3.01      & 0.19         & 4.59       & 0.29         & 5.66       & 0.35         & 5.98       & 0.37         \\ \hline
\end{tabular}
\caption{Performance Analysis of the Kernel Image Processing Algorithm}
\label{tab:table6}
\end{table}

\begin{minipage}{\textwidth}
\begin{lstlisting}[caption = Image Processing Network,label=lst:listing17]
int nodes = 1		// 1,2,4,8,16
String inFile = "./DSC_7610-1024.jpg"  //humming bird feeding from gladioli
String outFile = "./ DSC_7610-1024_${nodes}_Edge2.jpg"
//edge 1
Matrix kernel1 = new Matrix(rows: 3, columns: 3)
kernel1.entries = new int[3][3]
kernel1.setByRow([-1, -1, -1], 0)
kernel1.setByRow([-1,  8, -1], 1)
kernel1.setByRow([-1, -1, -1], 2)
//edge 2
Matrix kernel2 = new Matrix(rows: 5, columns: 5)
kernel2.entries = new int[5][5]
kernel2.setByRow([ -1, -1, -1,  -1, -1], 0)
kernel2.setByRow([ -1, -1, -1,  -1, -1], 1)
kernel2.setByRow([ -1, -1, 24,  -1, -1], 2)
kernel2.setByRow([ -1, -1, -1,  -1, -1], 3)
kernel2.setByRow([ -1, -1, -1,  -1, -1], 4)

def emitDetails = new DataDetails( dName: img.getName(),
                                   dInitMethod: img.initMethod,
                                   dCreateMethod: img.createMethod,
                                   dCreateData: [inFile, outFile])
def resultDetails = new ResultDetails( rName: imgRslt.getName(),
                                       rInitMethod: imgRslt.initMethod,
                                       rCollectMethod: imgRslt.collectMethod,
                                       rFinaliseMethod: imgRslt.finaliseMethod )

def emit = new Emit(eDetails: emitDetails)
def engine1 = new StencilEngine( nodes : nodes,
                               partitionMethod: img.partitionMethod,
                               functionMethod: img.greyScaleMethod )
def engine2 = new StencilEngine( nodes: nodes,
                               convolutionMethod: img.convolutionMethod,
                               convolutionData: [kernel2, 1, 0],
                               updateImageIndexMethod: img.updateImageIndex )
def collector = new Collect( rDetails: resultDetails)

\end{lstlisting}
\end{minipage}

\subsection{Goldbach Conjecture: Unstructured Data}

The Goldbach conjecture \cite{goldbach_conjecture} asserts that all even numbers greater than 2 can be expressed as the sum of two primes.  In this example, we initially generate primes, based on the method described in \cite{hansen-19}, up to just more than half the maximum Goldbach number required.  The primes are then used by several work processes that partition the Goldbach space, which determine that all the even numbers in that partition can be formed from the sum of two primes.  The final phase is to check each of the partitions to find out the maximum Goldbach number that has been found such that there is a continuous sequence from 2 up to that number.

\begin{figure}
        \centering
        \includegraphics[width=100mm]{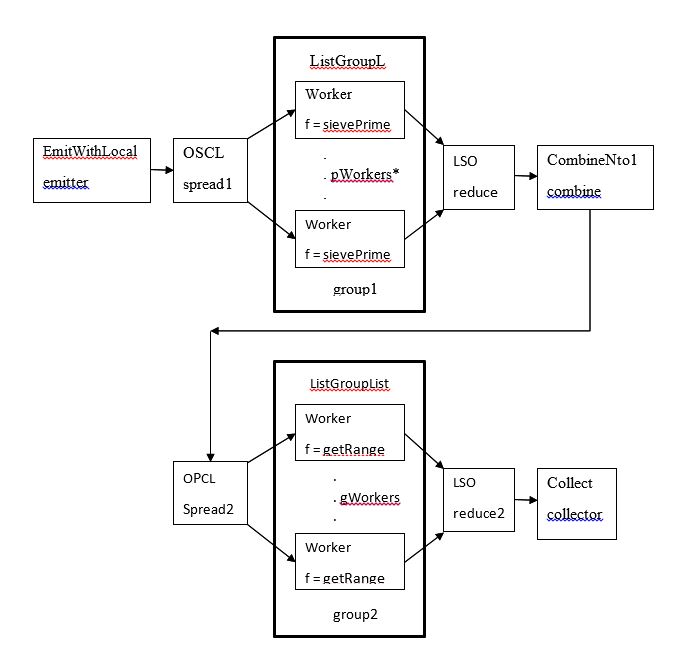}
        \caption{Goldbach Solution Architecture}
        \label{fig:fig6-2}
    \end{figure}

Figure \ref{fig:fig6-2} shows the process architecture and Listing \ref{lst:listing18} shows the structure of the network used to solve the problem .  It comprises two phases, the first determines the required prime numbers and the second part evaluates the Goldbach numbers.  The prime number phase uses an $EmitWithLocal$ process, which is like the previously discussed $Emit$ process but with the addition of an additional local class used during the data creation process.  The $prime$ object is emitted containing a single prime number.  The local $sieve$ class is used to obtain the next prime, up to the maximum prime specified by $filter$.  The emitted prime is then sent to a group of workers ($group1$) that partition the search space to find the multiples of the prime.  The emitted prime is transferred to all members of the group using a $OneSeqCastList$ spreader process.  In experiments it was found that the best value of $pWorkers$ is 1, that is a group of one worker.  This provides two processes finding the primes, the emit process finding the next prime and the group process(es) determining prime multiples in the whole search space.

The next stage is to combine all the partitions containing the prime numbers into one list using the $reduce1$ and $combine$ processes.  The $CombineNto1$ \{Listing \ref{lst:listing18}:22-24\} process inputs objects, until a $UniversalTerminator$ is read and is used to combine the input objects into a single output object.  The combined object, which contains all the primes is then sent to all $gWorkers$ \{Listing \ref{lst:listing18}:27\} the members of another group $group2$ using a $OneParCastList$ \{Listing \ref{lst:listing18}:25\} spreader process.  Each of these workers operates on an equal sized partition of the Goldbach space.  The method $getRange$  \{Listing \ref{lst:listing18}:30\} determines whether each number in the partition satisfies the Goldbach conjecture.  Finally, the output from each of the workers is sent to a $collector$ process using a $ListSeqOne$ reducing process $reduce2$ \{Listing \ref{lst:listing18}:31\}.  The $collector$ process determines the maximum number that has a Goldbach conjecture pair of prime numbers, which it prints.

\begin{minipage}{\textwidth}
\begin{lstlisting}[caption =Goldbach Conjecture Network,label=lst:listing18]
int maxPrime = 50000     // prime limit; Goldbach number < 100000
int pWorkers = 1    // number of prime workers
int gWorkers = 2    // number of Goldbach workers
int filter = Math.sqrt(maxPrime) + 1

. . . create the required details data objects

def eDetails = new DataDetails( dName:  prime.getName(),
                		  dInitMethod: prime.init,
                		  dCreateMethod: prime.create,
                		  lName: sieve.getName(),
                		  lInitMethod: sieve.init,
                		  lInitData: [filter])

def emitter = new EmitWithLocal(eDetails: eDetails)
def spread1 = new OneSeqCastList()
def group1 = new ListGroupList( gDetails: g1Details,
                                workers: pWorkers,
                                outData: false,
                                function: prime.sievePrime )
def reduce1 = new ListSeqOne ()
def combine = new CombineNto1( localDetails: combineLocal,
                               outDetails: combineOut,
                               combineMethod: internalList.toIntegers)
def spread2 = new OneParCastList()
def group2 = new ListGroupList ( gDetails: g2Details,
                                 workers: gWorkers,
                                 modifier:[[gWorkers], [gWorkers], [gWorkers], [gWorkers] ],
                                 outData: false,
                                 function: resultantPrimes.getRange)
def reduce2 = new ListSeqOne ( )
def collector = new Collect (rDetails: resDetails)
\end{lstlisting}
\end{minipage}

Table \ref{tab:table7} shows the speed up and efficiency of the Goldbach network for various values of MaxPrime.  In all cases the value of $pWorkers$ \{Listing \ref{lst:listing18}:18\} was set at 1.  As with previous examples the performance is limited by the number of available cores.

\begin{figure}
        \centering
        \includegraphics[width=80mm]{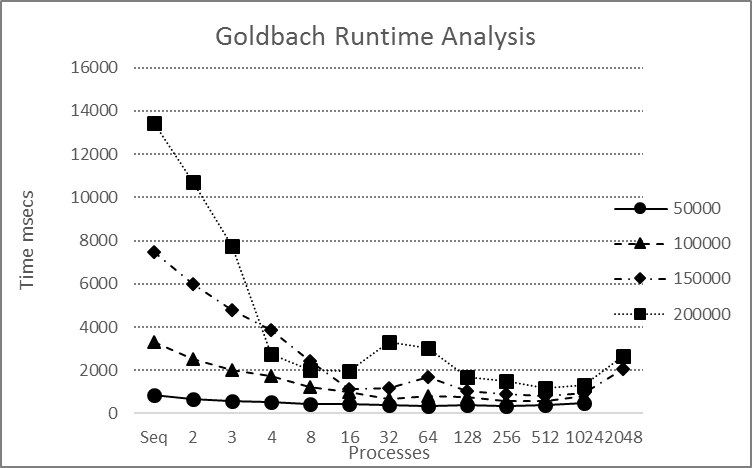}
        \caption{Runtime Analysis of the Goldbach Conjecture Algorithm}
        \label{fig:graph6}
    \end{figure}

Figure \ref{fig:graph6} shows the timing data for the Goldbach network and there are some unusual characteristics when the number of $gWorker$ \{Listing \ref{lst:listing18}:27\} processes increases beyond the number of available cores and hyper-threads.  This usually occurs because of interactions between data held in cache and bulk memory.  Were this to occur in an example where the cores were still being used then this would need to be further investigated, say by using the logging capability (Section 8). We note that when the maximum prime number is increased to 200000 the runtime increases when using 32 and 64 processes and then reduces when using 128-512 processes. This pattern is seen, to a lesser extent when max prime is set to 150000. This consistent increase in runtimes is not itself unexpected, but the subsequent reduction in runtime, prior to the expected increase when increasing to 2048 processes needs further investigation.  
 
 \begin{table}[]
\tiny
\centering
\begin{tabular}{|l|l|l|l|l|l|l|l|l|}
\hline
\textbf{Goldbach} & \textbf{Max Prime} & \textbf{50000}      & \textbf{Max Prime} & \textbf{100000}     & \textbf{Max Prime} & \textbf{150000}     & \textbf{Max Prime} & \textbf{200000}     \\ \hline
\textbf{gWorkers} & \textbf{SpeedUp}   & \textbf{Efficiency} & \textbf{SpeedUp}   & \textbf{Efficiency} & \textbf{SpeedUp}   & \textbf{Efficiency} & \textbf{SpeedUp}   & \textbf{Efficiency} \\ \hline
2                 & 1.28               & 64.20               & 1.31               & 65.30               & 1.25               & 62.51               & 1.26               & 62.85               \\ \hline
3                 & 1.45               & 48.49               & 1.64               & 54.70               & 1.56               & 52.17               & 1.74               & 57.91               \\ \hline
4                 & 1.63               & 40.82               & 1.91               & 47.78               & 1.92               & 48.07               & 4.88               & 121.93              \\ \hline
8                 & 1.98               & 24.69               & 2.75               & 34.36               & 3.09               & 38.68               & 6.73               & 84.10               \\ \hline
16                & 1.96               & 12.25               & 3.37               & 21.09               & 6.65               & 41.54               & 6.85               & 42.78               \\ \hline
32                & 2.09               & 6.53                & 4.99               & 15.60               & 6.22               & 19.44               & 4.04               & 12.62               \\ \hline
64                & 2.32               & 3.62                & 4.20               & 6.57                & 4.39               & 6.86                & 4.41               & 6.89                \\ \hline
128               & 2.13               & 1.66                & 4.37               & 3.41                & 7.22               & 5.64                & 8.04               & 6.28                \\ \hline
256               & 2.37               & 0.92                & 5.73               & 2.24                & 8.18               & 3.20                & 8.98               & 3.51                \\ \hline
512               & 2.30               & 0.45                & 5.53               & 1.08                & 9.26               & 1.81                & 11.32              & 2.21                \\ \hline
1024              & 1.79               & 0.18                & 4.00               & 0.39                & 7.95               & 0.78                & 10.04              & 0.98                \\ \hline
2048              &                    &                     &                    &                     & 3.64               & 0.18                & 5.02               & 0.25                \\ \hline
\end{tabular}
\caption{ Performance of the Goldbach Network (Empty cells indicate amount of work per worker too small)}
\label{tab:table7}
\end{table}
 
\subsection{Mandelbrot Set: Structured Grid}

The Mandelbrot Set is solved using a solution suggested in \cite{image_filtering} by using an escape value whereby once the number of iterations exceeds the escape value a colour of black is assumed.  The problem can be solved by either associating each pixel of the grid with a process or by processing a line of the grid.  The former is more appropriate for a GPGPU (General Purpose Graphics Processing Unit) based solution and the latter, adopted in this paper for a multi-core and cluster-based solution.  The architecture is a simple farm, using $any$ style connections meaning that as soon as one of the worker processes in $group$ \{Listing \ref{lst:listing19}:16\} becomes available it can process the next available line of the image.

\begin{minipage}[c]{0.95\textwidth}
\begin{lstlisting}[caption =Mandelbrot Network,label=lst:listing19]
int workers = 4                 // number of parallel worker cores
int maxIterations = 100         // number of iterations before escape
int width = 700                 //1400   700        350
int height = 400                //800    400        200
double pixelDelta = 0.005       //0.0025 0.005      0.01
def emitDetails = new DataDetails(dName: mandelbrotLine.getName(),
                                  dInitMethod: mandelbrotLine.init,
                                  dInitData: [width, height, pixelDelta, maxIterations],
                                  dCreateMethod: mandelbrotLine.create)
def resultDetails = new ResultDetails(rName: mandelbrotCollect.getName(),
                                      rInitMethod: mandelbrotCollect.init,
                                      rCollectMethod: mandelbrotCollect.collector,
                                      rFinaliseMethod: mandelbrotCollect.finalise)
def emit = new Emit(eDetails: emitDetails)
def spread = new OneFanAny(destinations: workers)
def group = new AnyGroupAny (function: mandelbrotLine.calcColour,
                    workers: workers)
def reduce = new AnyFanOne ( sources: workers)
def collector = new Collect(rDetails: resultDetails)
\end{lstlisting}
\end{minipage}

Table \ref{tab:table8} shows the speedup and efficiency obtained by the ‘embarrassingly parallel’ Mandelbrot solution.  Yet again performance is limited by the number of available cores but also shows that as the problem size increases by increasing the width, in pixels, of the image, the performance varies such that the speedup increases, and efficiency improves with problem size when the maximum number of cores are used.  It should be remembered that the other processes in the network shown in Listing \ref{lst:listing19} will also consume some resource and that these inconsistencies are observed when only half the number of cores are used.

\begin{table}[]
\tiny
\centering
\begin{tabular}{|l|l|l|l|l|l|l|}
\hline
\textbf{Mandelbrot} & \textbf{Width}   & \textbf{350}        & \textbf{Width}   & \textbf{700}        & \textbf{Width}   & \textbf{1400}       \\ \hline
\textbf{Processes}  & \textbf{SpeedUp} & \textbf{Efficiency} & \textbf{SpeedUp} & \textbf{Efficiency} & \textbf{SpeedUp} & \textbf{Efficiency} \\ \hline
1                   & 0.98             &                     & 1.01             &                     & 0.97             &                     \\ \hline
2                   & 1.69             & 84.53               & 1.87             & 93.65               & 1.75             & 87.43               \\ \hline
4                   & 2.30             & 57.58               & 2.77             & 69.16               & 2.89             & 72.29               \\ \hline
8                   & 2.26             & 28.29               & 2.78             & 34.79               & 2.88             & 36.05               \\ \hline
16                  & 2.30             & 14.35               & 2.76             & 17.23               & 2.89             & 18.05               \\ \hline
32                  & 2.30             & 7.19                & 2.75             & 8.60                & 2.91             & 9.09                \\ \hline
\end{tabular}
\caption{Mandelbrot Performance Analysis}
\label{tab:table8}
\end{table}

\begin{figure}
 
        \centering
        \includegraphics[width=80mm]{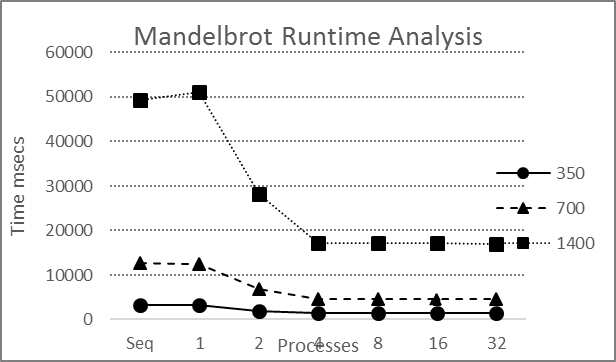}
        \caption{Runtime Analysis of the Mandelbrot Multi-core Algorithm  }
        \label{fig:graph7}
    \end{figure}

Figure \ref{fig:graph7} shows the runtime analysis of the Mandelbrot network, where the performance improvement achieved with problem size and number of cores (and hyper-threads) is more clearly observed.
 
\section{Utilising Workstation Clusters}
\label{sec:workstations}

The library contains a means of utilising workstation clusters.   One of the workstations is designated as the host node and the remainder as worker nodes.  The host node contains all the application object definitions and executes the emit and collect processes in the network.  A special set of cluster connectors are used to connect the emit process to the worker processes in each of the nodes, which use the Client-Server design pattern \cite{hansen_1973} .  Briefly, a client makes a request on a server and is then ready to receive any response from the server immediately.  A server on accepting a request from a client guarantees to respond within finite time.  If a server is unable to satisfy the response it may make a request as a client on another server.  Provided the network of clients and servers does not contain a loop then the network is guaranteed to be deadlock and livelock free and Welch \cite{welch_justo_wilcock_1993} provides a proof of this.

Each worker node has a process network that exploits the maximum number of available cores.  The host node builds objects that can be sent to each worker node which contain both the process definitions and the required network channel connections.  A network channel is one that uses the IP address for the connection.  Within the JCSP library the nature of a channel, be it internal or network, is transparent to the process definition.  Each worker node initially sends location information to the host, which can the construct the net channel connections for each worker node.  The host then sends the respective definitional object to each of the worker nodes that can then install the required processes.  A worker node runs a loader process that is independent of the node’s location or the process network to be installed.  The method is like that described in (\cite{kerridge_2014} pp. (ii)167-190).  Thus, the complete cluster can be initialised and run from a single host workstation.

The version of the Mandelbrot set used for these experiments was an image width of 5600 pixels and the escape value was 1000, substantially greater than that used for the multi-core version.  Table \ref{tab:table9} shows the speedup and efficiency obtained using various numbers of worker nodes.  As the number of worker nodes increase both the speedup and efficiency of the solution decrease but the network still achieves a creditable performance, only losing 0.01 when using a single cluster where the network comprises the host and one worker node using all its cores.

\begin{table}[h]
\tiny
\centering
\begin{tabular}{|l|l|l|}
\hline
\textbf{Nodes} & \textbf{Speedup} & \textbf{Efficiency} \\ \hline
1              & 0.99             & 0.99                \\ \hline
2              & 1.88             & 0.94                \\ \hline
3              & 2.73             & 0.91                \\ \hline
4              & 3.52             & 0.88                \\ \hline
5              & 4.13             & 0.83                \\ \hline
6              & 4.73             & 0.79                \\ \hline
\end{tabular}
\caption{Speedup of Mandelbrot Using Workstations}
\label{tab:table9}
\end{table}

   \begin{figure}[h]
        \centering
        \includegraphics[width=80mm]{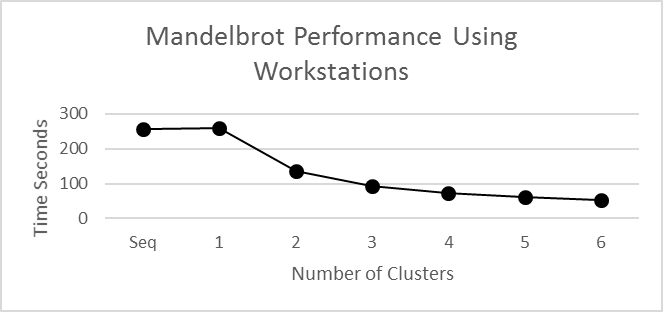}
        \caption{Runtime Analysis of the Cluster Solution for the Mandelbrot Algorithm}
        \label{fig:graph8}
    \end{figure}

Figure \ref{fig:graph8} shows the timing analysis and it can be observed that the performance of one cluster against a sequential (Seq) version is worse, due to the extra communication required.  Thereafter there is a performance improvement as the number of workstations is increased but this reduces as the number is increased further, with the required communication between the clusters and the emit and collector processes on the host starting to have an effect.  Increasing the problem size reduces this drop off to a higher number of workstations but is still present.  However, for this problem we can see a near linear speedup up to 4 nodes before the performance improvement starts to reduce.  
  
\section{Logging Process Networks}
The goal of any parallelization is to improve the performance of the solution.  The developer needs to know which parts of the algorithm are relatively slower than other parts.  For this we need to identify the time it takes to undertake specific parts of the algorithm.  To this end developers often instrument their applications to find out how long specific parts take.  Inevitably, this instrumentation is time-consuming and is often ad-hoc and application specific.  Conversely, many IDEs provide plugins that enable performance analysis but in general the plugins do not provide sufficiently detailed process level data.

The GPP library adopts a different approach.  Any terminal or functional process can invoke logging simply by giving the phase a name and the name of a property of the process’s input object that can be used to identify each object as they pass through the solution network.  Thus, logging can be undertaken on a per process basis or on a subset of the processes.  Logging is achieved by providing two versions of each process; first  a version with no logging and secondly, a version into which logging statements have been inserted.  This has the effect of enabling static compilation of the non-logged version and thereby the fastest execution speed.  The logged version uses reflection to obtain the value of the property that is being logged, thereby prohibiting the use of static compilation.  Log Messages are communicated to a Logging process which runs in parallel with the rest of the process network.  Logging is indicated by a special annotation for $gppBuilder$, which specifies the name of the file to which log output is written.  The resulting executable network has the required channel and process definitions automatically included.

Each log message includes an identifying tag together with a time, the name of the log phase and possibly the value of a property of the object that is being logged.  The logging messages are output to the console as well as being recorded in a file.  Thus, as the application network runs it is possible to see the messages appear and this gives a visual cue as to the parts of the network taking a long time to complete.

\subsection{Utilisation of Log Data – Concordance}

Applying the Logging capability to the Concordance system (See 6.1) it can be observed that the initial input and processing of the text file (Stage 1) for the ‘bible’ text consumes about 20\% of the total processing time.  It therefore makes sense to parallelise this stage as well.  The approach adopted is to read the file into blocks.  A block just contains unprocessed words.  Each block is then passed to another process which removes punctuation and calculates the integer value that represents each word.  The blocks are then recombined into a single data structure which is then emitted into the remainder of the network as described in Section 6.1.  The effect of this change was to reduce the total time taken by the application by at least 10\%, depending on the size of the text being analysed.

\section{Underpinning Theory and Network Refinement}
\subsection{Deadlock Analysis}

The GPP library uses the I/O-SEQ \cite{welch_justo_wilcock_1993} pattern for each of the processes defined in the library, apart from cluster communication processes.  The I/O-SEQ pattern has the form shown in Listing \ref{lst:listing20}.  It is assumed that at some point one of the compute elements sets $running$ to $false$  \{Listing \ref{lst:listing20}:5\} to terminate the loop.

\begin{minipage}{\textwidth}
\begin{lstlisting}[caption =I/O-SEQ Design Pattern,label=lst:listing20]
while (running) {
	... parallel inputs from all input channels
	... compute
... parallel outputs to all output channels
... compute
}
\end{lstlisting}
\end{minipage}

The use of the pattern can be seen in the coding for the Worker process (see Figure \ref{fig:3-1}) as shown in Listing \ref{lst:listing21}.

\begin{minipage}{\textwidth}
\begin{lstlisting}[caption =The main loop of the Worker process ,label=lst:listing21]
while (running){
    inputObject = input.read()				// read input object
    if ( inputObject instanceof UniversalTerminator){	// termination received
        running = false
        output.write(inputObject)
    }
    else {
        callUserMethod(inputObject, function, [dataModifier, wc], 1)
        output.write(inputObject)				// write modified input object
    }
\end{lstlisting}
\end{minipage}

The pattern is simply used because there is only one input and one output communication channel.  Also, there is only one compute phase represented by the call to the $callUserMethod()$ \{Listing \ref{lst:listing21}:8\}, which invokes the user defined method function on the $inputObject$ using the parameters $dataModifier$ and $wc$.  $DataModifer$ is a parameter passed to the process and $wc$ is a possibly null local worker class.  If the function fails the error message referred to as 1 will be output, together with a library defined message and the whole system will exit with a user defined negative error code.

\subsection{Network Refinement}

In addition to using CSPm \cite{gibson-robinson_armstrong_boulgakov_roscoe} and FDR \cite{gibson-robinson_armstrong_boulgakov_roscoe_2014}, the underlying JCSP library \cite{cspforjava} is based on the occam model \cite{inmos_ltd_1984} used to implement the CSP \cite{Hoare-1978} concepts.  Thus, we can use the Laws of occam \cite{roscoe_hoare_1988} to reason about equivalence between process networks in GPP. Listing \ref{lst:listing22} gives the representation of a parallel pipeline comprising processes P, Q and R, which are then replicated $pipes$ times.

\begin{minipage}{\textwidth}
\begin{lstlisting}[caption =Representation of a Parallel Group of Pipelines ,label=lst:listing22]
PAR i = 0 .. pipes
  PAR
    P
    Q
    R
    
\end{lstlisting}    
\end{minipage}

\begin{figure}
        \centering
        \includegraphics{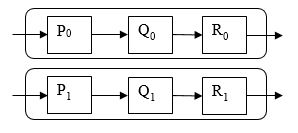}
        \caption{Expansion of Listing 22}
        \label{fig:fig9-1}
    \end{figure}

Expansion of  Listing \ref{lst:listing22} yields the structure shown in Figure \ref{fig:fig9-1} assuming $pipes$ has the value 1.
 
Listing \ref{lst:listing23} shows a different formulation which shows a parallel (PAR) of three replicated instances of P, Q and R processes.

\begin{minipage}{\textwidth}
\begin{lstlisting}[caption =Representation as a Group of Pipelines,label=lst:listing23]
PAR
  PAR i = 0 .. groups
    P
  PAR j = 0 .. groups
    Q
  PAR k = 0 .. groups
    R
\end{lstlisting}

\end{minipage}

\begin{figure}[h]
        \centering
        \includegraphics{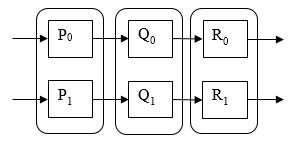}
        \caption{Expansion of Listing 23}
        \label{fig:fig9-2}
    \end{figure}

Similarly, expansion of Listing \ref{lst:listing23} yields the structure shown in Figure \ref{fig:fig9-2}, which by comparison with Figure \ref{fig:fig9-1} creates a similar structure assuming groups has the value 1.

Laws 5.2 and 5.3 of \cite{roscoe_hoare_1988} specify that a parallel (PAR) is associative and symmetric, provided all the channels used are private to each process.  Thus, we do not have to resort to diagrammatic representations to show equivalence we can use the laws already defined for occam.  Listings \ref{lst:listing22} and \ref{lst:listing23} expand to a $PAR (P_0, P_1, Q_0, Q_1, R_0, R_1)$.  Thus, by application of the laws we can create either of the two representations.  This means that any network created using GPP can be analysed and refined using well-developed theory as well as being checked for deadlock, divergence (livelock) and failures using CSPm and FDR as discussed in 4.6.

\section{Comparison with Other Approaches}

Fox et al \cite{fox_1988} describe three general features of parallel computation, assuming the algorithm is being applied to a large dataset. The features are described as:
\begin{itemize}
    \item being able to decompose the dataset into smaller elements
    \item large speedups can be achieved provided each processor has enough data
    \item each processor uses an algorithm like the sequential version, assuming each processor operates on part or all the dataset and boundary conditions mean that communication between processors is required.
\end{itemize}

The GPP library, implements these concepts directly and in addition allows the programmer to test a sequential version of the algorithm, without modification to any of the code by simply calling the required methods appropriately.

The state of the art up to 2010 is summarised in \cite{gonzalez_2010} and the overarching conclusion is that most skeletons resulted in the programmer having to contort their solution to fit the demands of the skeleton.  This made the algorithm unrecognisable as demonstrated in Listings 1 and 2 of the paper.  Table III of \cite{gonzalez_2010} has been updated and is still being maintained \cite{alg_skeleton} and shows that subsequent developments have attempted to make the programmer’s task simpler.  Intel \cite{intel} and Microsoft \cite{microsoft} have created libraries that assist in the creation of parallel systems based on underlying thread models.

OpenMP and MPI are commonly used parallelisation tools. OpenMP \cite{OpenMP} relies on the programmer adding directives and other control structures to an existing program and possibly adding additional subroutines to capture the parallel nature of an application, for example, adding a directive to parallelise a FOR statement. To do this the programmer has to fully understand the implications of the parallelisation and ensure the effect can be undertaken without creating access violations. In GPP the existing sequential code remains unaltered. MPI \cite{MPI} provides communication between nodes in a cluster using a large set of routines the user must incorporate into their code. Typically, these routines operate in pairs, one to send and one to receive a message. The user has several different choices over the style of communication, including synchronous, blocking, non-blocking and can combine different styles in the same communication. This makes it very hard for the user to reason about the behaviour of their program. In particular, the non-blocking routines allow a process to continue without having to wait for the sending or reading of a message and the programmer is required to test a buffer to determine whether a message has been communicated. GPP has just one form of communication, characterised as synchronous, blocking, and unbuffered that enable simple reasoning about communication patterns and can be formally analysed. GPP hides this complexity from the programmer.

A commonly used architecture is Map-Reduce \cite{mapreduce} which has been designed to filter and sort large quantities of distributed data obtained from various sources. The aim was also to insulate multi-processor hardware from variations in throughput over a network and the performance of different hardware components. Hadoop \cite{hadoop} provides an implementation of the architecture. A specialised distributed file store is created to which code is sent that filters and reduces data volumes before sending to another node for further processing. The Map-Reduce architecture is appropriate for processing large data volumes as occurs in ‘big data’ applications. GPP, in part, does contain this ability using cluster based computing and pipeline architectures, but it does not deal with node failures.

Leyton et al \cite{leyton_henrio_piquer_2010} investigated the problem of adding error handling into existing skeleton architectures using exception handling concepts found in many programming languages and discussed how these should be reported.  In GPP the approach adopted is that as soon as an error is found the system exits thereby making it easier for the programmer to determine the cause of the problem.  GPP also makes extensive use of assertions (\cite{Konig_2015} pp 31-46, 156-7 )to ensure that required parameters and values are present and valid before the parallel part of the processes are executed.  The FastFlow \cite{aldinucci_marco_kilpatrick_torquati_20111} development created a high-level data flow model for multi-core systems using C++ template libraries and is the first reference to emitter, workers and collectors, which were further refined in a significant development, RISC-pb2l, \cite{danelutto_torquati_2013}.  This introduced the idea of building blocks that could be composed into a variety of parallel architectures.  It introduced a mathematical notation but without the underpinning provided by CSP that lies at the core of GPP.  Aldinucci et al extended the RISC-pb2l \cite{aldinucci_campa_danelutto_kilpatrick_torquati_2013} to include further components to enable the construction of more complex parallel architectures, which was integrated into the FastFlow system.  Chalmers \cite{chalmers_2015} discusses the development of a process-based approach to skeletons and parallel design patterns arguing that this may be a more robust way of building parallel systems.  This approach was further extended in \cite{chalmers_jon_pederson_2016}, which used a prototype version of GPP to collect the results presented.
Danelutto et al in a guest editorial \cite{danelutto_pelagatti_torquati_2016} stated “it is crucial that the research community makes a significant progress toward making the development of parallel code accessible to all programmers”  and subsequent developments have shown that the concepts of Domain Specific Languages (DSL) have dominated \cite{danelutto_torquati_kilpatrick_2016} and \cite{griebler_danelutto_torquati_fernandes_2017}.  The advantage of GPP is that it is its own DSL with the advantage that much of the underlying detail is hidden from the programmer but with ability to add annotations that enable, for example, the logging capability.

\section{Evaluation}
At the outset of the project some goals were identified, each of which is now reviewed.
\subsection{ Extended collection of parallel building blocks}
The library comprises, terminal, functional and connector process types.  The functional processes that enable shared access data objects in a secure manner are novel in terms of their inclusion in a library.  This means that programmers used to a shared memory model of parallelism my find it easier to assimilate.  However, the library does not suffer from concurrent access to a data structure because the methods adopted in these processes specifically exclude such problems.  This detail is hidden from the user so that they need not even be aware of these problems.
\subsection{Formal proof of operation}
Unlike other such libraries, correctness of operation was designed in from the outset, initially using formally developed parallel design patterns.  Subsequently the correctness of combinations of processes was created using CSPm and FDR4.  The process design and implementation has ensured that each is built according to the proof that was developed for each type of process.  Further, we have developed proofs that show that the commonly used patterns Pipelines of Groups and Groups of Pipelines are equivalent to each other in that they refine each other.  Thus, the user can change the pattern knowing that the overall effect will be the same but the performance may be different.
\subsection{Integrated Logging}
The logging system is designed into the architecture.  It operates at the property level so user can monitor the movement of data objects as they pass through the process network, identifying possible bottlenecks.
\subsection{ DSL style and associated builder }
The library adopts a declarative style of specification, such that all the communication infrastructure between processes can be deduced and checked by  gppBuilder.  The user is only concerned with the ordering of the processes and the integration with the sequential parts of the application.  The builder will refuse to create a process network that does not ensure the correct communication structures between the processes.  Table \ref{tab:code} summarises the effect on code length that the DSL approach has achieved.

\begin{table}[]
\begin{tabular}{|c|c|c|c|}
\hline
Code Name                 & Listing Number  & Difference & \%   \\ \hline
Montecarlo (pattern)      & Listing 1 + 2   & 3          & 8\%  \\ \hline
Montecarlo(group)         & Listing 1 + 3   & 23         & 58\% \\ \hline
Montecarlo(pipeline)      & Fig 4           & 15         & 30\% \\ \hline
Concordance (PoG-pattern) & Listing 12 + 13 & 1          & 2\%  \\ \hline
Concordance (GoP-pattern) & Listing 12 + 14 & 3          & 6\%  \\ \hline
Jacobi                    & Listing 15      & 15         & 33\% \\ \hline
N-body                    & Listing 16      & 14         & 27\% \\ \hline
Image                     & Listing 17      & 19         & 30\% \\ \hline
Goldbach Conjecture       & Listing 18      & 30         & 42\% \\ \hline
\end{tabular}
\caption{Percentage Difference between the DSL specification and the Built code Difference is the number of additional lines added to the runnable built version \% expresses the difference as a percentage of the DSL specification}
\label{tab:code}
\end{table}

For those examples that use one of the available patterns the difference is less than 10\%.  For the remaining examples, the difference is substantially more.  This means  the user is only focusing on the process definitions and not having to be concerned with creating the communication structures.  If the application does not use logging, then the final executable code is as efficient as if it had been compiled by a Java compiler directly.  The use of the groovyJCSP library means the gppBuilder does not have to convert the DSL specification into Java.  From prior knowledge the use of groovyJCSP reduces the lines of code required by more than one-half.
\subsection{Use of extant sequential code}
The paper has not focussed on the sequential code because in most cases that code was either extant, N-body and Concordance or was based on pseudo code from websites.  The approach of a user providing method names rather than having to implement an interface was an explicit choice so that the method names in extant code do not have to be changed.  More importantly, because Groovy and Java components can be mixed the user can utilise Java code without alteration, as was the case in the N-Body solution.
\subsection{ Exploit parallelism easily and efficiently}
In all the examples we can see the effect of parallel processing is to improve performance.  It is also noticeable that the speedup for many of the applications is not very high when a larger number of processes are used.  The test machine had 4 cores and 4 hyper-threads thus in most of the examples we see improvement up to 4 and sometimes 8 processes but after that the performance is essentially flat.  The underlying processor has multiple cores but only accesses a single cache and memory.  It therefore must ensure, in hardware, that two cores are not accessing the same location at the same time which necessarily consumes resource.  The resulting program ensures that two processes cannot access the same memory location at the same time, but the hardware is unable to exploit this knowledge.
\subsection{Cluster based solutions}
The benefit of using the underlying JCSP library is that process definitions are not dependant on the nature of the communication channel, internal or networked.  Thus, the same process definitions can be used over a network.  This is particularly useful for the $connector$ processes which have the same definition regardless of communication method.  The way in which networked versions are invoked is different from internal communications, but the process body is identical.  The proofs developed previously still apply to a clustered solution.  The sophisticated type checking system used by Groovy makes this even simpler.

\section{Conclusions}
A major goal of the development of GPP was to make parallel programming as simple as possible.  The aim was to provide an environment whereby a programmer could take an extant sequential solution and parallelise it as easily as possible.  Most of the examples presented in this paper have taken published sequential solutions and incorporated them into a parallel solution.  The only one using a bespoke solution was the Goldbach (6.5) example but that is a more complex example and would only be attempted once the programmer has more confidence.  As Fox \cite{fox_1988}  suggests the naïve parallel programmer has to learn how to partition data sets.  GPP makes this as simple as possible by providing processes that enable the partitioning of large data sets in a controlled manner where the programmer just has to specify the size of the partitions and the data is then accessed in a safe manner without the programmer having to make further adjustments.

The choice of Groovy as the programming environment was crucial as it enabled the construction of the processes with simple strings for the method names.  This meant  the library processes could be constructed with a place-holder name for each method and the user just had to relate their method name with the one used in the process.  The ability to pass variable numbers of parameters to methods, as Lists was also crucial in making the coding simpler so the programmer is not limited to specific parameters that tend to constrain the other libraries discussed.

\section{Further Work}

Recently, a project has taken the log output and produced a visualisation of the running system, which enables the identification of possible processing bottlenecks.  Currently, the prototype visualisations are limited to several specific patterns.  This capability needs to be extended so that the visualisation can be deduced from the DSL specification of the application.  Currently, the library can use recursion only as part of a piece of sequential code.  We have been investigating how we can apply this to creating recursive processes so that a complete recursive process architecture can be built, initial results suggest that the concept has merit, provided the problem is sufficiently large.  We are also investigating how a specific architecture can be constructed that supports the parallelisation of evolutionary algorithms not by running the same algorithm as many different jobs, but by making the internal algorithm parallel by separating it into a server and multiple clients determining the fitness function for the subdivided population.
Finally, to support cluster based parallel computing we are investigating a builder that will create the required components to load into a cluster.  The code will be loaded to the cluster from a single host machine to all the nodes.  The mechanism will exploit the ability of the JCSP.net2 package to enable dynamic loading of object code on an as-needed basis. 

\section{Acknowledgements}
We acknowledge the many discussions we had with Kevin Chalmers and Jan Pedersen concerning the RISC-pb2l concepts that influenced an early version of the library.

\bibliographystyle{unsrt}  
\bibliography{refs}

\appendix

\section{Library Repository and Downloading Software}

All the software used in developing and subsequent use of the groovyParallelPatterns library has used the JetBrains Intellij development environment \url{https://www.jetbrains.com/idea/download}.  The development environment utilises Gradle build files \url{https://gradle.org/guides/}.  Every component in the Groovy Parallel Patterns ecosystem is supplied with the required Gradle build file.
The Groovy / Java environment used for development was Groovy-3 and the AdoptOpenJDK version 11, available at \url{https://adoptopenjdk.net/}.  The build files will need to be changed as later versions of Groovy are made available.  The version current during development was Groovy-3.0.7 available at \url{https://mvnrepository.com/artifact/org.codehaus.groovy/groovy-all}.

As Java and Groovy versions are released, users must download and install these separately and associate the Intellij project structure with that version.  Users of other IDEs will have to ensure they can build the software in their own environment.

Software is made available using the GitHub Package Repository mechanism.  Each Package contains the library jar, sources, pom and module files.  The libraries must be downloaded and extracted to the local Gradle repository.

The required libraries are automatically downloaded when the primary source download for all the demonstration software used in this paper is cloned.

The repository holding the demonstration examples is, \url{https://github.com/JonKerridge/GPP\_Demos} Release 1.1.12.  The repository README contains details of all the repositories and dependencies used in the build, so that users of other build tools can recreate the requirements easily.

Its Gradle build file will access all the other libraries used by the examples including the groovy\_parallel\_patterns library and its associated GPP\_Builder program.  In order to download Github Packages a user requires to have a Github Personal Access Token.

See \url{https://docs.github.com/en/github/authenticating-to-github/creating-a-personal-access-token}

A user will have to create their own \tt{gradle.properties} file at the same directory level as the \tt{build.gradle} file that contains

gpr.user=userName\\
gpr.key=userPersonalAccessToken\\

The library software also requires the JavaFX capability, used by the visualisation capability, which is downloaded as part of the build file for the Groovy Parallel Patterns library.  The version used is Open JFX 11 (\url{https://gluonhq.com/products/javafx/}), and thus the project must also use Java 11 or later.

\section{Prior Conference Presentation}
An early version of the library was presented at GR8Conf, 2016, Copenhagen.  A paper and a video of the presentation are available at:
\\
Paper: \\
\url{https://www.researchgate.net/publication/303913933_Groovy_Parallel_Patterns_A_Library_to_Support_Parallelization}
\\
Video:
\\
\url{https://www.youtube.com/watch?v=YtB8V37IS6k&feature=youtu.be&a}

\section{Test Machine Specifications}
\label{sec:appendixC}
The test PC comprises Intel Core i7-4790K, 4.00GHz processor with 16GB DDR3-1333 memory, running Windows 10 Pro.  The processor has 8MB cache, 4 cores plus 4 hyper threads.
\\
For the cluster-based examples the PCs were connected using a 1Gbit ethernet.

\end{document}